\documentclass[floatfix,twocolumn,showpacs,preprintnumbers,amsmath,amssymb,pra,superscriptaddress,longbibliography]{revtex4-1}
\usepackage{color}
\usepackage[usenames,dvipsnames,svgnames,table]{xcolor}
\usepackage[colorlinks=true,linkcolor=blue,urlcolor=blue,citecolor=blue]{hyperref}
\usepackage{mathtools}
\usepackage{graphicx}
\usepackage{dcolumn}
\usepackage{array}
\usepackage{lipsum}
\usepackage{bm}
\usepackage{subfigure}
\usepackage{amssymb}
\usepackage{multirow}
\usepackage{tabularx}
\usepackage{amsmath}
\usepackage{braket}
\graphicspath{{plots/}}
 \usepackage{lipsum}
\usepackage{mathrsfs}

\renewcommand{\vec}[1]{\mathbf{#1}}
\newcommand{\qv}{{\bf q}}

\begin{document}
\title{
Density Response of the Warm Dense Electron Gas beyond Linear Response Theory: Excitation of Harmonics
}

\author{Tobias Dornheim}
\email{t.dornheim@hzdr.de}

\affiliation{Center for Advanced Systems Understanding (CASUS), D-02826 G\"orlitz, Germany}

\affiliation{Helmholtz-Zentrum Dresden-Rossendorf (HZDR), D-01328 Dresden, Germany}

\author{Maximilian B\"ohme}
\affiliation{Center for Advanced Systems Understanding (CASUS), D-02826 G\"orlitz, Germany}
\affiliation{Helmholtz-Zentrum Dresden-Rossendorf (HZDR), D-01328 Dresden, Germany}

\affiliation{Technische  Universit\"at  Dresden,  D-01062  Dresden,  Germany}

\author{Zhandos A.~Moldabekov}

\affiliation{Center for Advanced Systems Understanding (CASUS), D-02826 G\"orlitz, Germany}
\affiliation{Helmholtz-Zentrum Dresden-Rossendorf (HZDR), D-01328 Dresden, Germany}

\author{Jan Vorberger}
\affiliation{Helmholtz-Zentrum Dresden-Rossendorf (HZDR), D-01328 Dresden, Germany}

\author{Michael Bonitz}

\affiliation{Institut f\"ur Theoretische Physik und Astrophysik, Christian-Albrechts-Universit\"at zu Kiel, D-24098 Kiel, Germany}

\begin{abstract}
In a recent Letter, Dornheim \textit{et al.}~[PRL \textbf{125}, 085001 (2020)] have investigated the nonlinear density response of the uniform electron gas in the warm dense matter regime. More specifically, they have studied the cubic response function at the first harmonic, which cannot be neglected in many situations of experimental relevance. In this work, we go one step further and study the full spectrum of excitations at the higher harmonics of the original perturbation based on extensive new \textit{ab initio} path integral Monte Carlo (PIMC) simulations. We find that the dominant contribution to the density response beyond linear response theory is given by the quadratic response function at the second harmonic in the moderately nonlinear regime. Furthermore, we show that the nonlinear density response is highly sensitive to exchange--correlation effects, which makes it a potentially valuable new tool of diagnostics. To this end, we present a new theoretical description of the nonlinear electronic density response based on the recent effective static approximation to the local field correction [PRL \textbf{125}, 235001 (2020)], which accurately reproduces our PIMC data with negligible computational cost.
\end{abstract}

\maketitle

\section{Introduction\label{sec:introduction}}

Linear response theory (LRT), i.e., the assumption of a linear response of a system of interest to a sufficiently small perturbation, is ubiquitous throughout physics and related disciplines. Prominent examples include the density and current response of many-body systems to external fields, e.g. \cite{quantum_theory}, the computation of optical absorption or plasmon spectra \cite{haug-jauho_book,kwong-etal.98pss}, the probing of superfluidity~\cite{cep,Shi_PRB_2005}, the description of electron-phonon interactions~\cite{PhysRevB.54.16487}, and dynamical simulations based on density functional theory~\cite{Gross_Kohn_PRL_1985}. 
From a practical point of view, the linear response only depends on the equilibrium properties of the unperturbed system, which often makes an accurate theoretical description feasible in the first place. This also allows one to characterize a system in terms of its response, which is of fundamental importance for diagnostics.

Yet, while it is clear that LRT is only accurate for small perturbations, often the validity of this assumption is not checked. On the one hand, this is quite understandable, as the general nonlinear response is substantially more complicated than LRT and, consequently, much poorer understood.
On the other hand, it is well known that nonlinear effects play an important role for many applications, e.g., in the optical excitation and ionization of atoms \cite{keldysh,schafer_prl_93}, in nonlinear optics \cite{blombergen}, in nuclear physics \cite{sakata}, in laser plasmas \cite{sprangle_prl_90}, in the inverse  bremsstrahlung heating of plasmas \cite{kremp_99_pre,bonitz_99_cpp}, nonlinear plasmons in magnetized plasmas \cite{bernstein,bonitz_prl_10,ott_pre_11}, the
excitation spectrum of graphene~\cite{PhysRevLett.105.097401,Mikhailov_PRL,Cox2016}, the optical properties of large molecules~\cite{PhysRevLett.67.1423}, or intersubband transitions in quantum-well semiconductor heterostructures~\cite{Lee2014}.

The present work is devoted to the nonlinear density response of warm dense matter (WDM), an exotic state with extreme temperatures ($T\sim10^4-10^8$K) and densities in the vicinity of solid states~\cite{wdm_book,fortov_review,new_POP}. These conditions occur naturally in astrophysical objects such as giant planet interiors~~\cite{Militzer_2008,militzer1} and brown dwarfs~\cite{saumon1,becker}. In addition, the compression path of a fuel capsule
towards inertial confinement fusion has to traverse the WDM regime~\cite{hu_ICF}. Finally, we mention the potential of hot electrons as a catalyst for chemical reactions~\cite{Brongersma2015}, and the recent discovery of new materials such as lonsdaleite~\cite{Kraus2016} and nanodiamonds~\cite{Kraus2017} at extreme conditions. Due to this gamut of applications, WDM has emerged as one of the most active frontiers in plasma physics and material science, and WDM is nowadays routinely realized in experiments at large research facilities around the globe; see the recent review article by Falk~\cite{falk_wdm} for an overview of different experimental techniques.

Unfortunately, the theoretical description of WDM is notoriously difficult due to the intriguingly intricate interplay of Coulomb correlations, thermal excitations, and quantum  effects. The correlation strength rules out perturbative expansions such as Green functions~\cite{kremp_book}, a non-vanishing temperature precludes the applications of the well-stocked arsenal of ground-state methods, and quantum effects make semi-classical methods like molecular dynamics insufficient. Therefore, computationally expensive quantum Monte Carlo methods~\cite{review,dornheim_pop17} have emerged as the method of choice to determine the basic properties of WDM~\cite{Driver_PRL_2012,Brown_PRL_2013,Militzer_PRL_2015,dornheim_prl,dornheim_prl_18,Malone_PRL_2016,groth_prl17}.

Like theoretical concepts in many other fields, current WDM theory is based, to a large extent, on LRT. For example, the theory of X-ray Thomson scattering~\cite{siegfried_review,kraus_xrts}, the most capable diagnostics in WDM experiments, presupposes a purely linear relation between perturbation and response. Other examples include the construction of electronically screened effective potentials~\cite{ceperley_potential,zhandos1,zhandos2}, the estimation of energy-relaxation rates~\cite{transfer1,transfer2,ernstorfer}, and the description of energy-loss properties like the stopping power~\cite{Cayzac2017,moldabekov_pre_20}.
In this context, the central property is often given by the linear density response function, $\chi(\mathbf{q},\omega)$ (with $\mathbf{q}$ and $\omega$ being the wave number and frequency of the perturbation), of the electrons in the system. These can be approximated by the uniform electron gas (UEG)~\cite{quantum_theory}, one of the most fundamental model systems in physics and quantum chemistry~\cite{loos,review}.
Consequently, many recent works~\cite{dornheim_pre17,groth_jcp17,dornheim_ML,hamann_prb_20,dynamic_folgepaper} have been devoted to the estimation of $\chi$, which has culminated in the complete characterization~\cite{dornheim_ML} of the static local field correction (LFC) $G(q)$ containing the full wave-number resolved information about electronic exchange--correlation effects (see also Eq.~(\ref{eq:LFC}) below). Furthermore, Dornheim~\textit{et al.}~\cite{Dornheim_PRL_2020_ESA,Dornheim_PRB_2021} have introduced the \emph{effective static approximation} to the LFC, which is available as a simple analytical representation~\cite{Dornheim_PRB_2021} and allows for the computation of highly accurate electronic properties within LRT with negligible computation cost.

In a recent Letter~\cite{Dornheim_PRL_2020}, three of us carried out extensive \textit{ab initio} path integral Monte Carlo (PIMC) simulations of the harmonically perturbed electron gas in the WDM regime without any assumptions about the response being linear. This allowed us to check the validity range of LRT, and we found that nonlinear effects cannot be neglected in many situations of experimental relevance. In addition, we used the exact PIMC data for the density response to extract the cubic density response function at the wave number of the perturbation itself.

In the present work, we substantially extend these efforts by considering the full excitation spectrum over the entire range of wave numbers. In particular, we find strong nonlinear excitations at the integer higher harmonics of the original perturbation. The quadratic response at the second harmonic constitutes the largest nonlinear contribution for moderate perturbations. Furthermore, we find that the nonlinear response functions strongly depend on electronic exchange--correlation effects, thus making them a potentially valuable new tool of diagnostics. To this end, we introduce a new theory of the nonlinear density response based on available representations of the LFC~\cite{dornheim_ML,Dornheim_PRB_2021}, which is capable to accurately reproduce our PIMC data with negligible
computational cost.

The paper is organized as follows: In Sec.~\ref{sec:theory}, we introduce the relevant theoretical background, including the PIMC method (\ref{sec:pimc}), the simulation idea and setup (\ref{sec:simulation_idea}), the extended theory of the nonlinear density response functions (\ref{sec:nonlinear_theory}), and the specific way how the latter can be estimated from our PIMC data (\ref{sec:approach}).
In Sec.~\ref{sec:results}, we present our extensive new simulation results covering a metallic density (\ref{sec:metallic}), the temperature dependence (\ref{sec:temperature}), and a strongly coupled case (\ref{sec:strong_coupling}).
The paper is concluded by a brief summary and outlook in Sec.~\ref{sec:summary}.

\section{Theory\label{sec:theory}}

We assume Hartree atomic units throughout this work. From a theoretical perspective, the WDM regime is typically defined by two characteristic parameters that are both of the order of one: (i) the density parameter (Wigner-Seitz radius) $r_s=\overline{r}/a_\textnormal{B}$, where $\overline{r}$ and $a_\textnormal{B}$ are the average inter-particle distance and first Bohr radius, and (ii) the reduced temperature $\theta=k_\textnormal{B}T/E_\textnormal{F}$, with $E_\textnormal{F}$ being the usual Fermi energy~\cite{quantum_theory}. A third principle parameter is given by the spin-polarization $\xi=(N^\uparrow-N^\downarrow)/N$. Throughout this work, we restrict ourselves to the case of a purely unpolarized system, i.e., $N^\uparrow=N^\downarrow=N/2$ and $\xi=0$.

\subsection{Path Integral Monte Carlo\label{sec:pimc}}

We carry out path-integral Monte-Carlo (PIMC) simulations~\cite{cep,Berne_JCP_1982,Takahashi_Imada_PIMC_1984} of $N$ electrons in the canonical ensemble, i.e., the volume $V=L^3$ (with $L$ being the box length), inverse temperature $\beta=1/k_\textnormal{B}T$, and particle number $N$ are constant. 

The basic idea of the PIMC method is to stochastically sample the thermal density matrix
\begin{eqnarray}
\rho(\mathbf{R},\mathbf{R},\beta) = \bra{\mathbf{R}}e^{-\beta\hat H}\ket{\mathbf{R}}\ ,
\end{eqnarray}
where $\mathbf{R}=(\mathbf{r}_1,\dots,\mathbf{r}_N)^T$ contains the coordinates of all $N$ electrons. In the end, the partition function is re-cast into the integral over $P$ density matrices at a $P$ times higher temperature, and the resulting high-dimensional integrals are evaluated using the Metropolis Monte-Carlo method~\cite{metropolis} to avoid the curse of dimensionality of standard quadrature methods; see Ref.~\cite{cep} for an extensive review article on the PIMC method.

An additional obstacle regarding the PIMC simulations of electrons is the requirement of anti-symmetry under particle exchange, which renders the partition function into a sum over both positive and negative contributions. The resulting cancellation of terms with opposite sign is the origin of the notorious fermion sign problem~\cite{troyer,dornheim_sign_problem} that leads to an exponential increase in computation time both with increasing the system size $N$ and with decreasing temperature $T$~\cite{dornheim_sign_problem}.

For this reason, one often employs approximate methods like the restricted path-integral Monte-Carlo method~\cite{Ceperley1991} that are not afflicted with the sign problem. Unfortunately, this advantage comes at the cost of an in practice uncontrolled systematic error~\cite{schoof_prl15,dornheim_prl16,Malone_PRL_2016}. Therefore, we here avoid such nodal restrictions and carry out the computationally expensive, yet exact standard PIMC simulations and deal with the sign problem by increasing the computation time, where this is feasible.

For completeness, we mention that all PIMC results presented in this work have been obtained using a canonical adaption~\cite{mezza} of the worm algorithm by Boninsegni \textit{et al.}~\cite{boninsegni1,boninsegni2}.

\subsection{Simulation idea\label{sec:simulation_idea}}
Following Refs.~\cite{dornheim_prl_20,bowen2,moroni,moroni2,groth_jcp,dornheim_pre}, we simulate a uniform electron gas that is subject to an external static harmonic perturbation. The corresponding Hamiltonian is given by
\begin{eqnarray}\label{eq:hamiltonian}
\hat H = \hat H_\textnormal{UEG} + 2 A \sum_{l=1}^N \textnormal{cos}\left( \hat{\mathbf{r}}_l\cdot{\mathbf{q}} \right)\ ,
\end{eqnarray}
where $\hat H_\textnormal{UEG}$ is the standard (unperturbed) UEG Hamiltonian~\cite{loos,quantum_theory,review}. The second term on the RHS. of Eq.~(\ref{eq:hamiltonian}) is the external potential corresponding to a single cosine of wave vector $\mathbf{q} = 2\pi/L (n_x,n_y,n_z)^T$ (with $n_i\in\mathbb{Z}$, and $L$ being the length of the simulation box) and the perturbation amplitude $A$. As $A$ corresponds to an energy in this notation, it is given in Hartree throughout this work. Moreover, we restrict ourselves to a strictly static perturbation as PIMC simulations of a time-dependent system are severely hampered by an additional dynamic sign problem~\cite{PhysRevLett.100.176403}.

To measure the density response, we compute the induced density
\begin{eqnarray}\label{eq:rho}
\braket{\hat\rho_\mathbf{k}}_{q,A} = \frac{1}{V} \left< \sum_{l=1}^N e^{-i\mathbf{k}\cdot\hat{\mathbf{r}}_l} \right>_{q,A} \ , 
\end{eqnarray}
where $\braket{\dots}_{q,A}$ indicates the parameters of the perturbation in Eq.~(\ref{eq:hamiltonian}), and $\mathbf{k}$ corresponds to the wave vector at which we measure the response of the system.
In particular, within linear response theory it holds
\begin{eqnarray}\label{eq:LRT}
\braket{\hat\rho_\mathbf{k}}_{q,A} = \delta_{\mathbf{q},\mathbf{k}} \chi^{(1)}(\mathbf{q}) A\ ,
\end{eqnarray}
with $\chi^{(1)}(\mathbf{q})$ being the standard linear response function~\cite{quantum_theory}. The latter is conveniently expressed as
\begin{eqnarray}\label{eq:LFC}
\chi^{(1)}(q) = \frac{\chi^{(1)}_0(q)}{1-\frac{4\pi}{q^2}\left[ 1-G(q)\right]\chi^{(1)}_0(q)}\ ,
\end{eqnarray}
where $v(q)=4\pi/q^2$ is the Fourier transform of the Coulomb potential (we use atomic units throughout), $\chi^{(1)}_0(q)$ is the density response function of the ideal (non-interacting) Fermi gas that is known from the literature~\cite{quantum_theory}, and $G(q)$ is the static local field correction containing the full wave-number resolved information about exchange--correlation effects. In particular, highly accurate data for $G(q)$ have recently become available as a neural-net representation that was constructed on the basis of extensive PIMC simulations~\cite{dornheim_ML,Dornheim_PRL_2020_ESA}.

In addition, we mention that the full wave-number dependence of
$\chi^{(1)}(q)$ can be computed from a single simulation of the unperturbed UEG via the imaginary-time version of the fluctuation--dissipation theorem, which states that
\begin{eqnarray}\label{eq:static_chi}
\chi^{(1)}(\mathbf{q}) = -n\int_0^\beta \textnormal{d}\tau\ F(\mathbf{q},\tau) \quad ,
\end{eqnarray}
with $F(q,\tau)$ being the intermediate scattering function~\cite{siegfried_review} evaluated at an imaginary time argument $\tau\in[0,\beta]$, see Refs.~\cite{dornheim_ML,dornheim_dynamic,dynamic_folgepaper,Dornheim_Vorberger_finite_size_2020,Boninsegni_maximum_entropy,Vitali_PRB_2010,Saccani_Supersolid_PRL_2012,Boninsegni1996,Motta_JCP_2015} for different applications of this quantity.

Finally, we mention the density of the perturbed electron gas in coordinate space, which, in LRT, is given by
\begin{eqnarray}\label{eq:coordinate_response_LRT}
n(\mathbf{r}) = n_0 + 2A \textnormal{cos}\left(
\mathbf{q}\cdot\mathbf{r}
\right)\chi^{(1)}(\mathbf{q}) \ ,
\end{eqnarray}
with $n_0$ being the average value of the density.

\subsection{Theory of nonlinear density response\label{sec:nonlinear_theory}}

The nonlinear density response of plasmas and condensed matter systems in and out of equilibrium has been studied in some detail theoretically and by simulations, e.g.,  \cite{Langreth1976,golden_pra_85,Bonitz_PRB_1994,kwong_prl-00}. Here we concentrate on the nonlinear response functions in thermodynamic equilibrium.
The general definitions of the second and third order response functions are given by an expansion of the induced density, $n_{\rm ind}(\vec r)=n(\vec r)-n_{0}$, according to a weak perturbation assumption~\cite{PhysRevB.37.9268,Bergara1999}%
 \begin{align}
n_{\rm ind}(\vec r)\! &=\!\!\int\!\mathrm{d} \vec r^{\prime} \chi(\vec r, \vec r^{\prime}) V(\vec r^{\prime}) \nonumber\\
&+\int\! \mathrm{d} \vec r^{\prime} \mathrm{d} \vec r^{\prime \prime} \mathscr{Y}(\vec r,\vec r^{\prime},\vec r^{\prime \prime})V(\vec r^{\prime})V(\vec r^{\prime \prime}) \nonumber \\
&+\int\! \mathrm{d} \vec r^{\prime} \mathrm{d} \vec r^{\prime \prime} \mathrm{d} \vec r^{\prime \prime \prime} \mathscr{Z}(\vec r,\vec r^{\prime},\vec r^{\prime \prime}, \vec r^{\prime \prime \prime})V(\vec r^{\prime})V(\vec r^{\prime \prime})V(\vec r^{\prime  \prime \prime}) \nonumber\\
&+\cdots,
\label{eq:n_ind_re}
\end{align}
where $V$ is the perturbing  potential,  whereas $\mathscr{Y}$ and $\mathscr{Z}$ are the second order and third  order response functions, respectively. 
%
%

Taking into account that, for a homogeneous system, $\chi(\vec r, \vec r^{\prime})=\chi(\vec r- \vec r^{\prime})$, $\mathscr{Y}(\vec r,\vec r^{\prime},\vec r^{\prime \prime})= \mathscr{Y}(\vec r-\vec r^{\prime},\vec r-\vec r^{\prime \prime})$, and $\mathscr{Z}(\vec r,\vec r^{\prime},\vec r^{\prime \prime}, \vec r^{\prime \prime \prime})=\mathscr{Z}(\vec r-\vec r^{\prime}, \vec r-\vec r^{\prime \prime}, \vec r - \vec r^{\prime \prime \prime})$,  Eq.~(\ref{eq:n_ind_re}) can be rewritten in Fourier space as ($\Omega$ denotes the volume)
\begin{align}
n_{\rm ind}(\vec k)\! &=\! \chi(\vec k) V(\vec k)  \nonumber\\
& + \frac{1}{\Omega}\sum_{\vec k_2} \! \mathscr{Y}(\vec k-\vec k_2,\vec k_2)V(\vec k-\vec k_2)V(\vec k_2)\nonumber\\
&+\frac{1}{\Omega^2}\sum_{\vec k_2}\sum_{\vec k_3}\, \mathscr{Z}(\vec k-\vec k_2+\vec k_3,\vec k_2,\vec k_3) \nonumber\\
&\times V(\vec k-\vec k_2+\vec k_3)V(\vec k_2)V(\vec k_3) +\cdots,
\label{eq:n_ind_q}
\end{align}
where wave-vector notation $\vec k$ is used to avoid confusion with the wave-vector $\vec q$ of the external harmonic field  $V(\vec r)=2A\cos{\vec q\cdot \vec r}$. 


By substituting into Eq.~(\ref{eq:n_ind_q}) the Fourier representation of the external potential ($A=\widetilde A/\Omega$), 
\begin{equation}
V(\vec k)= \widetilde A \left[\delta_{\vec k,\vec q}+\delta_{\vec k,-\vec q}\right], 
\end{equation}
we arrive at
\begin{align}
   n_{\rm ind}(\vec k)\! &=\! \widetilde A\left[\chi^{(1)}(\vec k)\delta_{\vec k,\vec q}+ \chi^{(1)}(\vec k)\delta_{\vec k,-\vec q}\right] \nonumber \\
   &+ \frac{\widetilde A^2}{\Omega} \Big[\mathscr{Y} (\vec k-\vec q,\vec q)\delta_{\vec k,2\vec q} 
   +\mathscr{Y}(\vec k+\vec q,-\vec q)\delta_{\vec k,-2\vec q}\Big] \nonumber\\
   &+ \frac{\widetilde A^3}{\Omega^2}\Big[\chi^{(1,\textnormal{cubic})}(\vec k,\vec q)\delta_{\vec k,\vec q}
   +\chi^{(1,\textnormal{cubic})}(\vec k,-\vec q)\delta_{\vec k,-\vec q}\nonumber\\
&\quad\qquad +\mathscr{Z}(\vec k-2\vec q,\vec q,\vec q)\delta_{\vec k,3\vec q}\nonumber\\
&\quad\qquad +\mathscr{Z}(\vec k+2\vec q,-\vec q,-\vec q)\delta_{\vec k,-3\vec q}\Big]\cdots.
\label{eq:n_ind_q2}
\end{align}
The first line contains the linear response $\chi^{(1)}$ at the incoming wave vector $\qv$. The second line collects the quadratic response $\mathscr{Y}$ at the second harmonic $2\qv$. The third line gives the cubic response at the first harmonic $\chi^{(1,\textnormal{cubic})}$
where the following notation for brevity was introduced 
\begin{align}\label{eq:tilde_chi}
   \chi^{(1,\textnormal{cubic})}(\vec k, \vec k^{\prime})&=\mathscr{Z}(\vec k, \vec k^{\prime},-\vec k^{\prime})+    \mathscr{Z}(\vec k, -\vec k^{\prime},\vec k^{\prime})\nonumber\\
    &+\mathscr{Z}(\vec k-2\vec k^{\prime}, \vec k^{\prime},\vec k^{\prime}).
\end{align}
Finally, the third and fourth lines contain the cubic response function $\mathscr{Z}$ at the third harmonic $3\qv$. We will drop all negative wave vectors from now on as the behavior is symmetric for positive and negative $\qv$. It should also be noted that only integer multiples of the original wave vector appear as possible higher order excitations.

 Now, by performing an inverse Fourier transform of Eq.~(\ref{eq:n_ind_q2}), we obtain, for the induced density in real space,
\begin{align}\label{eq:rho12}
n_{\rm ind}(\mathbf{r}) &=\frac{1}{\Omega}\sum_{\mathbf{k}} n_{\rm ind}(\vec k) \exp(i\vec k \vec r) \nonumber\\
                        &=2 \sum_{\eta=1}^\infty \braket{\hat\rho_{\eta\mathbf{q}}}_{q,A} \textnormal{cos}\left(\eta\mathbf{q}\cdot\mathbf{r} \right)
\end{align}
where, for $\eta=1$, $\eta=2$, and $\eta=3$:
\begin{eqnarray}\label{eq:rho1}
\braket{\hat\rho_\mathbf{q}}_{q,A} &=& \chi^{(1)}(q) A + \chi^{(1,\textnormal{cubic})}(q) A^3\ ,\\
\label{eq:rho2}
\braket{\hat\rho_\mathbf{2q}}_{q,A} &=& \chi^{(2)}(q) A^2 \ , \\
\label{eq:rho3}
\braket{\hat\rho_\mathbf{3q}}_{q,A} &=& \chi^{(3)}(q) A^3 \ ,
\end{eqnarray}
and, using Eqs.~(\ref{eq:n_ind_q2}) and (\ref{eq:tilde_chi}) we identify
\begin{align}
  \chi^{(2)}(q)&=\mathscr{Y}(\vec k-\vec q,\vec q)\delta_{\vec k,2\vec q}, \label{eq:chi2_Y}\\
  \chi^{(1,\textnormal{cubic})}(q)&=\Big[\mathscr{Z}(\vec q, \vec q,-\vec q)+\mathscr{Z}(\vec q, -\vec q,\vec q) \nonumber\\
    &\qquad+\mathscr{Z}(\vec k-2\vec q, \vec q,\vec q)\Big]\delta_{\vec k,\vec q}\,,\label{eq:chi1_Z}\\ 
  \chi^{(3)}(q)&=\mathscr{Z}(\vec k-2\vec q,\vec q,\vec q)\delta_{\vec k,3\vec q}.\label{eq:chi3_Z}  
\end{align}
%

Equations~(\ref{eq:chi2_Y}) and (\ref{eq:chi3_Z}) show the connection between the general second order, $\mathscr{Y}$, and third order, $\mathscr{Z}$, response functions with the response functions of the system at the second and third harmonics [$\chi^{(2)}(q)$ and $\chi^{(3)}(q)$, respectively]. Furthermore, Eq.~(\ref{eq:chi1_Z}) shows the connection of the cubic response at the first harmonic, $\chi^{(1,{\rm cubic})}(q)$, to the third order response function $\mathscr{Z}$.\\ 

 By definition, the total density in coordinate space $n(\mathbf{r})$ can as well be written as the Fourier series over all wave numbers $\mathbf{q}$,
 Yet, as only the harmonics of the perturbation wave vector $\mathbf{q}$ exhibit a nonzero response, the density is simply
 \begin{eqnarray}\label{eq:expansion}
 n(\mathbf{r}) = n_0 + 2 \sum_{\eta=1}^\infty
 \braket{\hat\rho_{\eta\mathbf{q}}}_{q,A}
 \textnormal{cos}\left(
 \eta\mathbf{q}\cdot\mathbf{r}
 \right)\ .
 \end{eqnarray}


\subsubsection{Noninteracting case}

A general expression for the ideal second order response function, $\mathscr{Y}_{0}$, and some of its properties were considered in Refs. \cite{PhysRevB.59.10145, CENNI1988279, PhysRevB.56.15654, PhysRevB.37.9268, Kalman, Rennert, Paasch, PhysRevB.52.13883, PhysRevE.54.3518, PhysRevA.11.2147, PhysRevA.17.390}.

Mikhailov~\cite{Mikhailov_Annalen,Mikhailov_PRL} expressed the ideal response functions at the second harmonic, $\chi^{(2)}_0(q)$, and third harmonic, $\chi^{(3)}_0(q)$, in terms of the first order response function, $\chi_0^{(1)}(q)$, without invoking $\mathscr{Y}_{0}(\vec k, \vec q)$ and $\mathscr{Z}_{0}(\vec k, \vec k^{\prime}, \vec q)$, but by directly expanding the induced density at harmonics. 
The recursion relations derived by Mikhailov~\cite{Mikhailov_Annalen,Mikhailov_PRL} are
\begin{eqnarray}\label{eq:Mikhailov2}
\chi^{(2)}_0(q) = \frac{2}{q^2}\left( \chi^{(1)}_0(2q)-\chi^{(1)}_0(q)\right),
\end{eqnarray}
and \cite{Mikhailov_PRL}
\begin{eqnarray}\label{eq:Mikhailov3}
\chi_0^{(3)}(q)=\frac{3\chi_0^{(1)}(3q)-8\chi^{(1)}_0(2q)+5\chi_0^{(1)}(q)}{3q^4}.
\end{eqnarray}
Such a recursion formula does not exist for the ideal cubic response at the first harmonic $\chi^{(1,\textnormal{cubic})}(q)$. In fact, the terms making up the ideal cubic response at the first harmonic diverge.


\subsubsection{Random Phase Approximation}

Taking into account screening in linear approximation, the RPA result for the response function from the Green's functions based consideration has the following form \cite{PhysRevB.37.9268}
\begin{align}\label{eq:Y_RPA}
    \mathscr{Y}(\vec k-\vec q, \vec q)&=\frac{\mathscr{Y}(\vec k-\vec q, \vec q)}{ \left[1-v(\vec q)\chi^{(1)}_{0}(\vec q)\right]^{2}\left[1-v(\vec k)\chi_0^{(1)}(\vec k) \right]}\,.
\end{align}
At the second harmonic, $\chi^{(2)}_{\rm RPA}( q)$ is found using the relation  (\ref{eq:chi2_Y}) 
and Eq.~(\ref{eq:Y_RPA})
\begin{equation}\label{eq:chi2_RPA}
    \chi^{(2)}_{\rm RPA}( q)= \frac{\chi^{(2)}_{0}( q)}{\left[1-v(q)\chi^{(1)}_{0}(q)\right]^{2} \left[1-v(2q)\chi^{(1)}_{0}( 2q)\right]}.
\end{equation}

To illustrate the level of approximation in Eqs.~(\ref{eq:Y_RPA}) \& (\ref{eq:chi2_RPA}), it is instructive to provide an alternative derivation of Eq.~(\ref{eq:chi2_RPA}).
To begin with, we recall that, in contrast to the ideal response function, the response function in RPA takes into account inter-electronic interactions in a mean field approximation. Therefore, introducing a total potential as a sum of the external potential and the Hartree potential due to the induced electronic density, $\Phi_{\rm tot}=\Phi_{\rm ext}+\Phi_{\rm ind}$, the induced density in RPA at the second harmonic is obtained in terms of the ideal response functions
\begin{align}\label{eq:n_ind2_RPA}
n_{\rm ind}^{\rm RPA}(2q)&=\chi_{0}^{(2)}(q) \left[\Phi_{\rm tot}(q)\right]^2+\chi_{0}^{(1)}(2q) \Phi_{\rm ind}(2q)\nonumber\\
&+\chi_{0}^{(1, cubic)}(2q) \left[\Phi_{\rm ind}(2q)\right]^3,
\end{align}
where the first term on the r.h.s. of Eq.~(\ref{eq:n_ind2_RPA}) is the second order response of the system which is proportional to the square of the perturbing potential [as can be seen from Eq.(\ref{eq:rho2})],  the second and third terms on the r.h.s. of Eq.~(\ref{eq:n_ind2_RPA}) are contributions from the linear and cubic responses induced by $\Phi_{\rm ind}$ at $2q$ [in accordance with Eq.~(\ref{eq:rho1})]. 

On the other hand, the induced density in RPA is expressed as the response to the external field using the quadratic response function in RPA
\begin{eqnarray}\label{eq:n_ind2_RPA2}
n_{\rm ind}^{\rm RPA}(2q)=\chi_{\rm RPA}^{(2)} (q) \left[\Phi_{\rm ext}(q)\right]^2.
\end{eqnarray}
From  Eqs.~(\ref{eq:n_ind2_RPA2}) and (\ref{eq:n_ind2_RPA}), we find
\begin{align}\label{eq:chi_gen_2}
    \chi_{\rm RPA}^{(2)} (q) \left[\Phi_{\rm ext}(q)\right]^2&=\chi_{0}^{(2)}(q) \left[\Phi_{\rm tot}(q)\right]^2+\chi_{0}^{(1)}(2q) \Phi_{\rm ind}(2q)\nonumber\\
&+\chi_{0}^{(1, cubic)}(2q) \left[\Phi_{\rm ind}(2q)\right]^3.
\end{align}
To solve Eq.~(\ref{eq:chi_gen_2}), we combine the Poisson equation for the induced potential at the second harmonic with Eq.~(\ref{eq:n_ind2_RPA2}) and find a relation between $\Phi_{\rm ind}(2q)$ and $\Phi_{\rm ext}(q)$
\begin{align}\label{eq:poisson}
    \Phi_{\rm ind}(2q)=\Phi_{\rm tot}(2q)&=\frac{4\pi}{(2q)^2} n_{\rm ind}^{\rm RPA}(2q)\nonumber\\
    &=\frac{4\pi}{(2q)^2}\chi_{\rm RPA}^{(2)} (q) \left[\Phi_{\rm ext}(q)\right]^2,
\end{align}
where it was taken into account that there is no contribution from the external potential to the total potential at the second harmonic. 

Secondly, we neglect the contribution from the cubic term, i.e. set $\chi_{0}^{(1, cubic)}(2q)=0$ in Eqs.~(\ref{eq:n_ind2_RPA}) \& (\ref{eq:chi_gen_2}), and, thus, approximate the total potential at the first harmonic using linear response theory, which reads in RPA
\begin{equation}\label{eq:phi_total}
    \Phi_{\rm tot}(q)=\frac{\Phi_{\rm ext}(q)}{1-v(q)\chi^{(1)}_{0}(q)}.
\end{equation}
Using Eqs.~(\ref{eq:poisson}) and (\ref{eq:phi_total}), we can solve Eq. (\ref{eq:chi_gen_2}) with respect to $\chi_{\rm RPA}^{(2)} (q)$  to find Eq. (\ref{eq:chi2_RPA}). 

For further discussion of the QMC data, it is important to emphasize  that, in $\chi_{\rm RPA}^{(2)}$ defined by  Eq. (\ref{eq:chi2_RPA}), (i) the screening is taken into account only in linear approximation and (ii) the contribution due to the cubic  response function is neglected.   

Next, by following the steps from Eq.~(\ref{eq:n_ind2_RPA}) to Eq.~(\ref{eq:chi_gen_2}), but for the induced density at the third harmonic, and making use of the screening in linear approximation, we find for the response function at the third harmonic in RPA
\begin{equation}\label{eq:chi3_RPA}
    \chi^{(3)}_{\rm RPA}( q)= \frac{\chi^{(3)}_{0}( q)}{\left[1-v(q)\chi^{(1)}_{0}(q)\right]^{3} \left[1-v(3q)\chi^{(1)}_{0}( 3q)\right]}.
\end{equation}
Note that, for a weak perturbation, the neglected contribution to the screening due to the cubic response, $\chi_0^{(1,\textnormal{cubic})}$, in Eq.~(\ref{eq:chi2_RPA}) is a higher order correction compared to the quadratic response. In contrast, for $ \chi^{(3)}_{\rm RPA}$, the neglected cubic response $\chi_0^{(1,\textnormal{cubic})}$  has the same order contribution as $ \chi_0^{(3)}$. Therefore, it can be expected that Eq.~(\ref{eq:chi2_RPA}) performs better than Eq. (\ref{eq:chi3_RPA}) when applied for the description of the QMC data at the corresponding harmonics.

The corresponding cubic response in RPA at the first harmonic is given by
\begin{equation}\label{eq:cubic_first_RPA}
    \chi^{(1,\textnormal{cubic})}_{\rm RPA}( q)= \frac{\chi^{(1,\textnormal{cubic})}_{0}( q)}{\left[1-v(q)\chi^{(1)}_{0}(q)\right]^{4}}.
\end{equation}
It is known that there are further terms that are technically contributing to the cubic response at the level of RPA. These terms are made up from non-diagonal quadratic response functions entirely~\cite{Rostami_2017,Vorberger_2021}.

\subsubsection{Going beyond the RPA description by using LFC }

In Eqs.~(\ref{eq:chi2_RPA}) and (\ref{eq:chi3_RPA}), the electronic interactions effect is included on the basis of the linear response functions in the denominator. Therefore, in analogy to the usual practice in linear response theory, we can go beyond RPA by introducing a local field correction (LFC), $G(q)$.  In this way, we arrive at the following equations for the response functions at the second and third harmonics with higher order electronic exchange-correlations effect included by using LFC 
 \begin{eqnarray}\label{eq:quadratic_LFC}
     \chi^{(2)}_{\rm LFC}( q) &=&  \chi^{(2)}_{0}( q) \left[1-v(q)\left[1-G(q)\right]\chi^{(1)}_{0}(q)\right]^{-2}\\  & & \times \left[1-v(2q)\left[1-G(2q)\right]\chi^{(1)}_{0}( 2q)\right]^{-1}, \label{eq:chi2_LFC}
 \end{eqnarray}
and 
 \begin{eqnarray}\nonumber
     \chi^{(3)}_{\rm LFC}( q) &=&  \chi^{(3)}_{0}( q) \left[1-v(q)\left[1-G(q)\right]\chi^{(1)}_{0}(q)\right]^{-3}\\  & & \times \left[1-v(3q)\left[1-G(3q)\right]\chi^{(1)}_{0}( 3q)\right]^{-1}.\label{eq:chi3_LFC}
 \end{eqnarray}
Equations~(\ref{eq:chi2_LFC}) and (\ref{eq:chi3_LFC}) provide an improved description of the response functions on the second and third harmonics with electronic exchange-correlations effect taken into account on the level of the linear response theory. 
We still give the respective result for the cubic response at the first harmonic
\begin{equation}\label{eq:cubic_first_LFC}
    \chi^{(1,\textnormal{cubic})}_{\rm RPA}( q)= \frac{\chi^{(1,\textnormal{cubic})}_{0}( q)}{\left[1-v(q)\left[1-G(q)\right]\chi^{(1)}_{0}(q)\right]^{4}}.
\end{equation}
The temperature and density dependent static LFC has been obtained from QMC simulations recently ~\cite{dornheim_pre17,groth_jcp17,dornheim_ML,hamann_prb_20,dynamic_folgepaper,dornheim_ML}. Furthermore, Dornheim~\textit{et al.}~\cite{Dornheim_PRL_2020_ESA,Dornheim_PRB_2021} have introduced the new concept of the \emph{effective static approximation} to the LFC, which is available as a simple analytical representation~\cite{Dornheim_PRB_2021}.

\subsection{\label{sec:approach}PIMC approach to nonlinear density response functions}

To extract the nonlinear density response functions of different harmonics from our PIMC simulation, we evaluate Eq.~(\ref{eq:rho}) given above for different perturbation amplitudes $A$ and subsequently perform fits to these data.

In particular, the density response in reciprocal space for the same wave number as the perturbation is fitted to~\cite{moroni,moroni2}
\begin{eqnarray}\label{eq:fit}
\braket{\hat\rho_\mathbf{q}}_{q,A} = \chi^{(1)}(q) A + \chi^{(1,\textnormal{cubic})}(q) A^3\ ,
\end{eqnarray}
where the coefficients $\chi^{(1)}(q)$ and $\chi^{(1,\textnormal{cubic})}(q)$ are the free parameters. Note that the determination of $\chi^{(1)}(q)$ from the PIMC data via Eq.~(\ref{eq:fit}) is redundant as it is already known from previous simulations of the unperturbed UEG via Eq.~(\ref{eq:static_chi}). However, these independent benchmark data can be compared to our new results for $\chi^{(1)}(q)$ and, thus, constitute a valuable consistency check of our approach, see also Ref.~\cite{dornheim_prl_20}. Furthermore, it is important to note that Eq.~(\ref{eq:fit}) only holds up to a maximum perturbation strength, beyond which contributions with a higher order of $A$ start to substantially contribute; see Sec.~\ref{sec:results} for a hands-on discussion of this point.

The density response of the second harmonic is, in first order in $A$, given by
\begin{eqnarray}\label{eq:second_fit}
\braket{\hat\rho_{2\mathbf{q}}}_{q,A} = \chi^{(2)}(q)A^2\ ,
\end{eqnarray}
where $\chi^{(2)}(q)$ is the free parameter that is obtained from a fit to the $A$-dependence for a fixed perturbation wave number $q$. Similarly, the density response of the third harmonic is obtained by fitting the PIMC data for the response at $\mathbf{k}=3\mathbf{q}$ to
\begin{eqnarray}\label{eq:third_fit}
\braket{\hat\rho_{3\mathbf{q}}}_{q,A} = \chi^{(3)}(q)A^3\ .
\end{eqnarray}

\section{Results\label{sec:results}}

\subsection{\label{sec:metallic}Metallic Density: $r_s=2$}

\begin{figure}\centering
\includegraphics[width=0.485\textwidth]{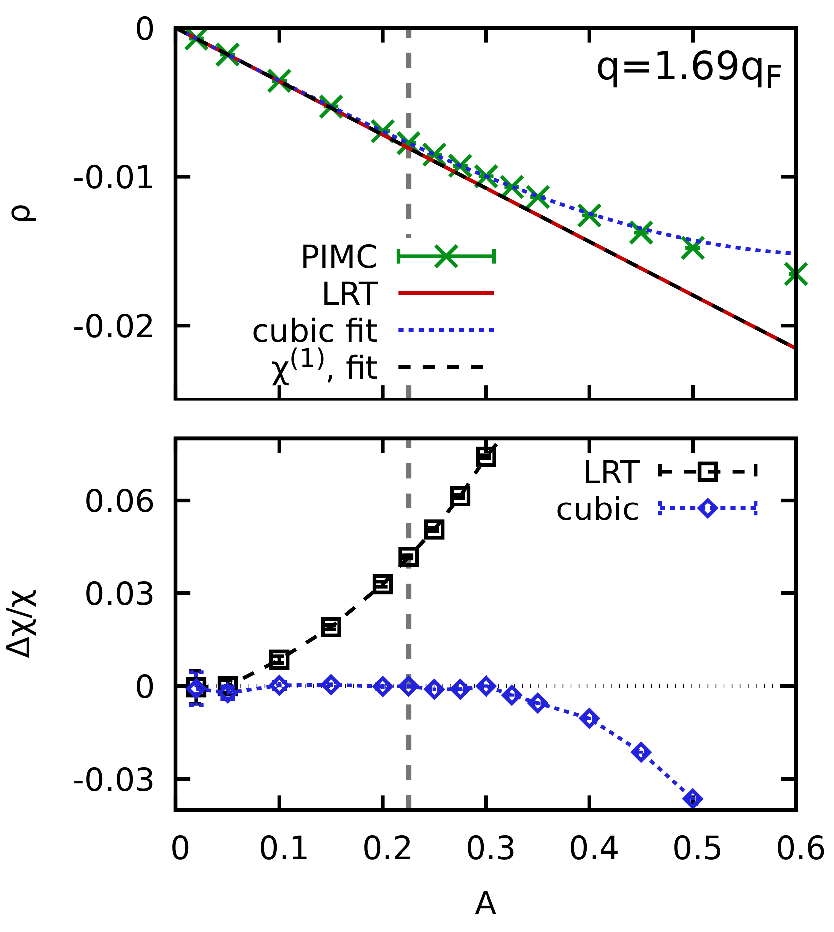}
\caption{\label{fig:PRL_rs2_theta1}
Dependence of the density response of the first harmonic $\braket{\hat\rho_\mathbf{q}}_{q,A}$ of the UEG on the perturbation amplitude $A$ for $N=14$, $r_s=2$, and $\theta=1$ with the perturbation wave number $q\approx1.69q_\textnormal{F}$. The green crosses are PIMC data evaluated from Eq.~(\ref{eq:rho}), and the solid red line is the prediction from LRT, cf.~Eq.~(\ref{eq:static_chi}). The dotted blue line has been obtained by fitting Eq.~(\ref{eq:fit}) to the PIMC data within the interval $A\in[0,0.225]$ (vertical grey dashed line), and the dashed black line shows the corresponding LRT prediction from the first coefficient only. The bottom panel shows the deviation of LRT and the cubic fit to the PIMC data.
}
\end{figure}

Let us start our investigation by considering the response of the electron gas at $r_s=2$ and $\theta=1$. These conditions are located at the center of the WDM regime and are realized experimentally for example in experiments with aluminum, e.g., Ref.~\cite{Sperling_PRL_2015}. In Fig.~\ref{fig:PRL_rs2_theta1}, we show the density response of the system in dependence on the perturbation amplitude $A$ for the same wave number as the perturbation, i.e., for the first harmonic $\braket{\hat\rho_\mathbf{q}}_{q,A}$. The green crosses show the raw PIMC data which have been obtained by evaluating Eq.~(\ref{eq:rho}). The solid red line shows the prediction by LRT and has been obtained from a previous simulation of the unperturbed UEG~\cite{dornheim_ML} by evaluating Eq.~(\ref{eq:static_chi}). Evidently, LRT is in excellent agreement with the PIMC data for small $A$ only and the disagreement rapidly increases with $A$. This can be seen particularly well in the bottom panel of Fig.~\ref{fig:PRL_rs2_theta1} showing the relative deviation to the PIMC data. For completeness, we mention that a similar analysis has been presented in Ref.~\cite{Dornheim_PRL_2020}, although for a different wave number.

\begin{figure*}\centering
\includegraphics[width=0.95\textwidth]{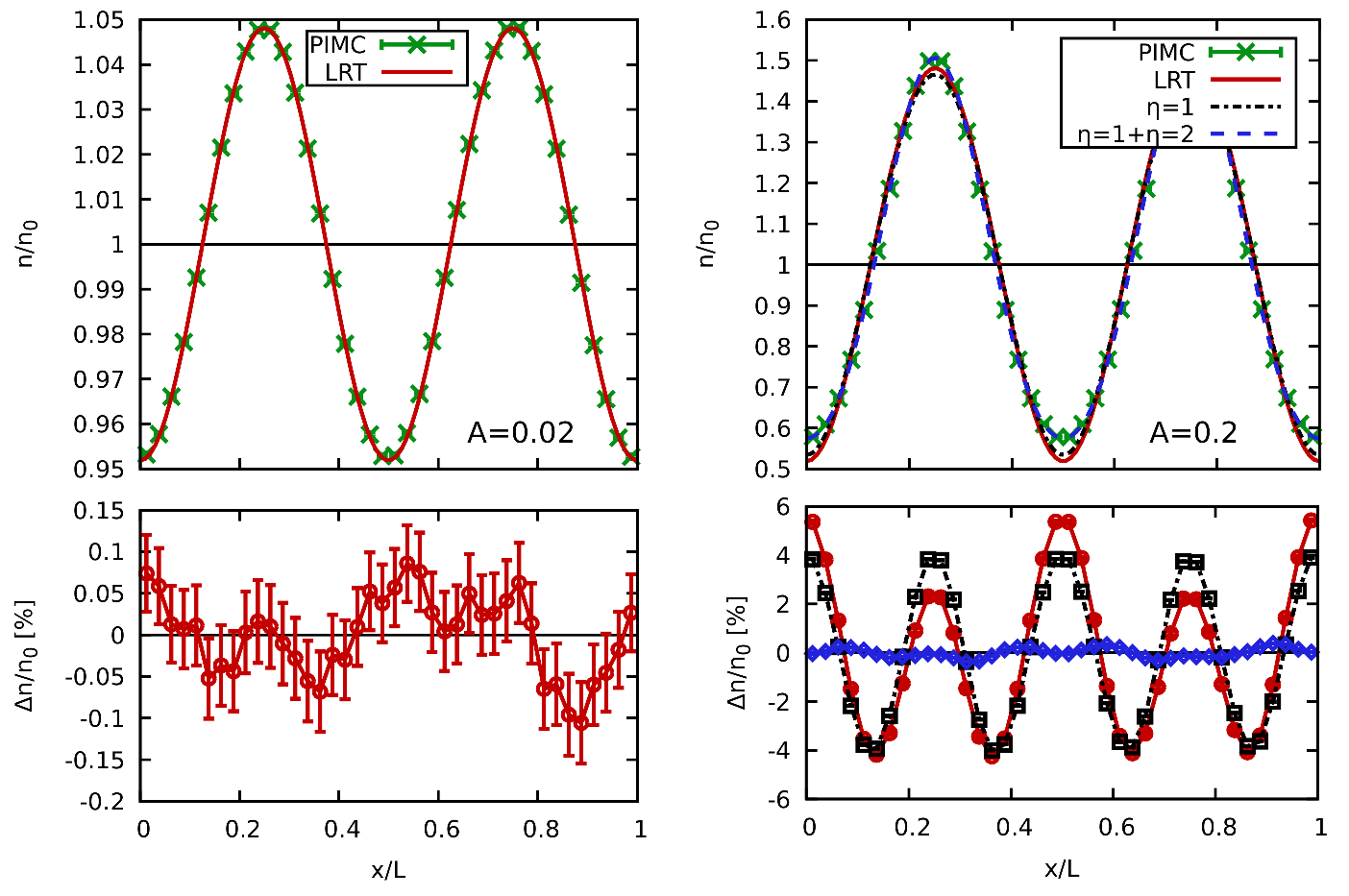}
\caption{\label{fig:density_qx2}
Density profile of the harmonically perturbed electron gas for $N=14$, $r_s=2$, and $\theta=1$ with $q\approx1.69q_\textnormal{F}$ and $\mathbf{q}=2\pi/L(2,0,0)^T$ for $A=0.02$ (left) and $A=0.2$ (right). Top row: density $n(x)$ along the direction of the perturbation, with the green crosses and solid red curves corresponding to PIMC data and LRT [cf.~Eq.~(\ref{eq:coordinate_response_LRT})]. The dash-dotted black and dashed blue curves have been obtained by truncating Eq.~(\ref{eq:expansion}) after the first ($\eta=1$) and second ($\eta=2$) harmonic, respectively.
The bottom panels show the corresponding deviation of different theories to the PIMC data.
}
\end{figure*} 

The dotted blue curve has been obtained by fitting Eq.~(\ref{eq:fit}) to the PIMC data within the interval $A\in[0,0.225]$, see also the vertical dashed grey line. Evidently, this curve captures the emerging deviations between the green crosses and the red line and remains accurate for substantially larger values of $A$. In particular, it remains accurate even beyond the fitting interval, which constitutes a strong empirical confirmation of the functional form in Eq.~(\ref{eq:fit}). In addition, the dashed black line shows the LRT prediction using the first coefficient from the fit and perfectly agrees with the independent red line. Still, we note that the blue curve, too, eventually becomes inaccurate, as the higher-order terms in $A$ that are neglected in Eq.~(\ref{eq:fit}) start to significantly contribute.

A possibly more intuitive illustration of the impact of the harmonic perturbation on the system is presented in Fig.~\ref{fig:density_qx2} where we show the density of the system along the direction of the perturbation. More specifically, the left column corresponds to $A=0.02$, which falls well into the LRT regime, cf.~Fig.~\ref{fig:PRL_rs2_theta1}. Consequently, the red curve that has been obtained from Eq.~(\ref{eq:coordinate_response_LRT}) is in excellent agreement to the green crosses depicting the PIMC data. The relative deviation of LRT to the latter is shown in the bottom panel of the same figure, and we find perfect agreement within the given Monte-Carlo error bars that are of the order of $\Delta n/n\lesssim0.1\%$. Overall, we find that the density modulation attains an amplitude of almost $5\%$ of the unperturbed density.

Let us next turn to the right column of Fig.~\ref{fig:density_qx2} showing the same quantity for a larger perturbation amplitude, $A=0.2$. In this case, the density modulation is substantially larger and the amplitudes of $n(x)$ are of the order of $50\%$ with respect to the unperturbed density. Thus, the system is indeed to a large degree shaped by the presence of the external potential. Consequently, the red LRT curve only gives a qualitative description of the density and systematically underestimates (overestimates) the height of the peaks (depth of the minima) of $n$. This can again be seen most clearly in the deviation plot in the bottom panel, where we find differences between LRT and the PIMC data exceeding $5\%$.

\begin{figure*}\centering
\includegraphics[width=0.97\textwidth]{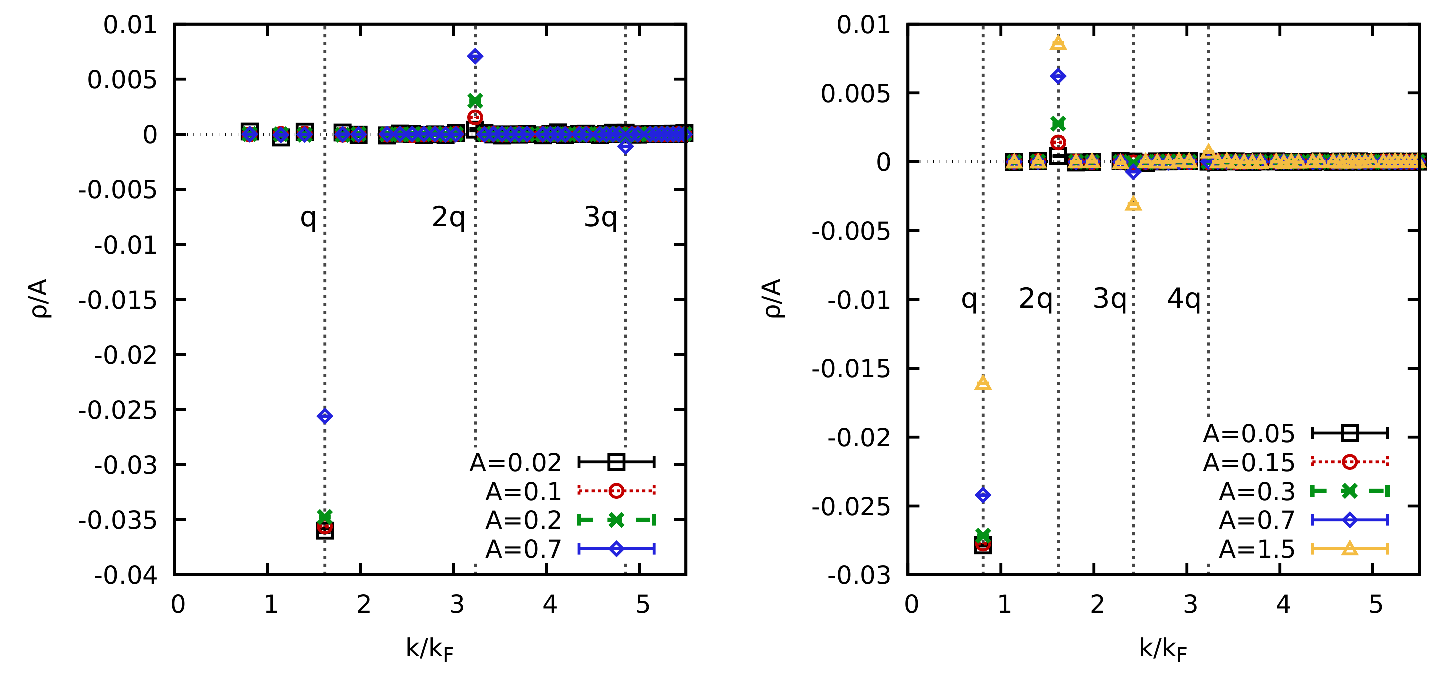}
\caption{\label{fig:spectrum_rs2}
Full wave-number dependence of the density response $\braket{\hat\rho_\mathbf{k}}_{q,A}$ of the UEG for $N=14$, $r_s=2$, and $\theta=1$ with $q\approx1.69q_\textnormal{F}$ (left) and $q\approx0.84q_\textnormal{F}$ (right) for different values of the perturbation amplitude $A$. The vertical dotted lines indicate the location of the first harmonics.
}
\end{figure*}

As a first step towards an improved description of the density profile $n(x)$, we go beyond LRT by including the cubic contribution $\chi^{(1,\textnormal{cubic})}$ to the first harmonic. Indeed, Fig.~\ref{fig:PRL_rs2_theta1} clearly indicates that this provides a fully adequate description of $\braket{\hat\rho_\mathbf{q}}_{q,A}$, and the dash-dotted black curve in Fig.~\ref{fig:density_qx2} corresponds to Eq.~(\ref{eq:expansion}) truncated after $\eta=1$. Let us list a few observations: i) the impact of $\chi^{(1,\textnormal{cubic})}$ on $n(x)$ is relatively small despite the large density modulation amplitude and the inaccurate description of LRT; ii) from the top panel, we see that $\chi^{(1,\textnormal{cubic})}$ leads to a somewhat improved agreement with the PIMC data around the minima, but increased deviations around the peaks.
This becomes more clear by looking at $\Delta n/n_0$ in the bottom panel. The LRT curve exhibits a deviation profile that oscillates twice as fast in space as the original density modulation, but the amplitude of this deviation is not constant. In fact, this deviation profile is a combination of a) the insufficient description of $\braket{\hat\rho_\mathbf{q}}_{q,A}$ by LRT (cf.~Fig.~\ref{fig:PRL_rs2_theta1}) and b) the omission of the second harmonic that does, by definition, oscillate twice as fast in space as the first one. By including $\chi^{(1,\textnormal{cubic})}$ in the description of $\braket{\hat\rho_\mathbf{q}}_{q,A}$, we have removed effect a), and the residual deviation profile (black squares in the bottom panel of Fig.~\ref{fig:density_qx2}) does indeed have a constant amplitude and corresponds to the contribution of the second harmonic.

Evaluating Eq.~(\ref{eq:expansion}) up to $\eta=2$ gives the dashed blue curve, which is in striking agreement with the green crosses, over the entire $x$-range. This is again confirmed by the deviation plot in the bottom, where the corresponding blue diamonds fluctuate around zero deviation with a relative accuracy of $\Delta n\sim0.1\%$.

A more systematic investigation of the emerging impact of higher harmonics upon increasing the perturbation amplitude is presented in Fig.~\ref{fig:spectrum_rs2}, where we show the full wave-number dependence of $\braket{\hat\rho_\mathbf{k}}_{q,A}$ for different values of $A$. The left panel shows results for $q\approx1.69q_\textnormal{F}$, i.e., the same perturbation wave number as in Figs.~\ref{fig:PRL_rs2_theta1} and \ref{fig:density_qx2}. First and foremost, we empirically confirm that a non-zero value for the density response can indeed only be found at the integer harmonics of the original perturbation wave vector $\mathbf{q}$. For the smallest depicted value of the perturbation amplitude, $A=0.02$ (black squares), we only find a signal at $q$ itself, which agrees with the prediction known from LRT. Increasing the perturbation strength by a factor of five ($A=0.1$) leads to the results given by the red circles. For the first harmonic, the response remains almost unchanged and varies only by $\sim1\%$ (cf.~the bottom panel of Fig.~\ref{fig:PRL_rs2_theta1}). At the same time, we find a significant response at the second harmonic that cannot be neglected and substantially contributes to observables like the density profile in coordinate space $n(\mathbf{r})$ that was discussed above.

A further increase of $A$ by a factor of two is depicted by the green crosses. In this case, the response at the first harmonic clearly deviates from the prediction by LRT, and we have already seen that the cubic contribution $\chi^{(1,\textnormal{cubic})}(q)$ is required for an adequate description. Further, we observe an even more pronounced deviation from LRT at the second harmonic, which was shown to be responsible for the bulk of the deviation between LRT and the PIMC data for $n(\mathbf{r})$ in Fig.~\ref{fig:density_qx2}. At the same time, no significant response is observed for higher harmonics ($\eta>2$) at these conditions.

This changes for the largest depicted perturbation strength shown in Fig.~\ref{fig:spectrum_rs2}, i.e., $A=0.7$ (blue diamonds). In this case, the signal at the first harmonic is substantially reduced compared to the other data points, and even the cubic response function $\chi^{(1,\textnormal{cubic})}(q)$ cannot provide a reasonable description of the density response at these conditions, see Fig.~\ref{fig:PRL_rs2_theta1}. Moreover, we find a large signal at the second harmonic, as it is expected, and a significant contribution for the third harmonic, $k=3q$.

\begin{figure}\centering
\includegraphics[width=0.475\textwidth]{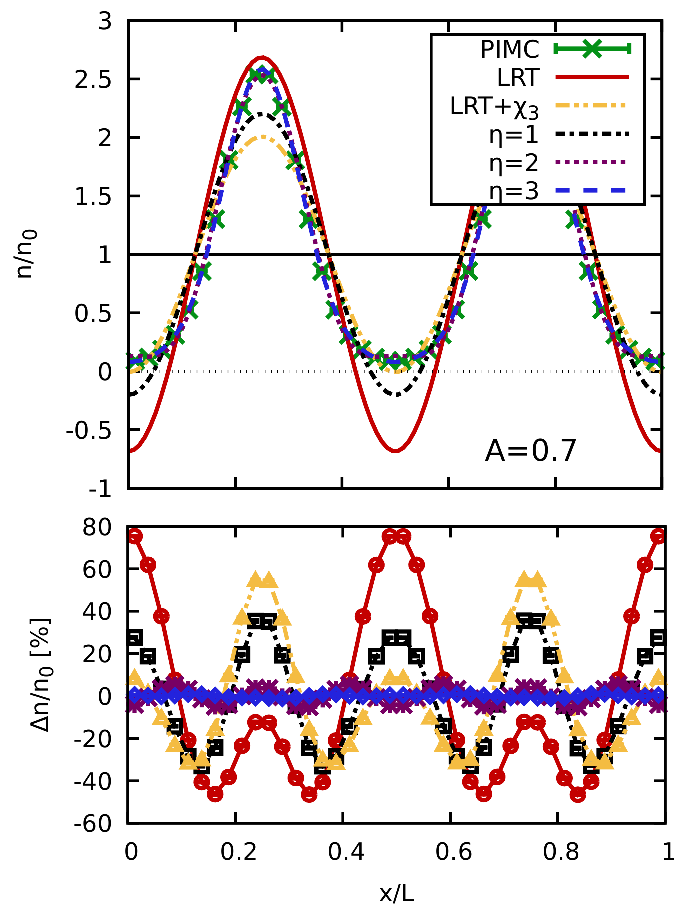}
\caption{\label{fig:density_qx2_A0p7}
Density profile of the harmonically perturbed electron gas for $N=14$, $r_s=2$, and $\theta=1$ with $q\approx1.69q_\textnormal{F}$ and $\mathbf{q}=2\pi/L(2,0,0)^T$ for $A=0.7$. Top panel: density $n(x)$ along the direction of the perturbation, with the green crosses and solid red curves corresponding to PIMC data and LRT [cf.~Eq.~(\ref{eq:coordinate_response_LRT})], and the yellow dash-double-dotted, black dash-dotted, purple dotted, and blue dashed line corresponding to different evaluations of Eq.~(\ref{eq:expansion}), see the main text.
The bottom panels show the corresponding deviation of different theories to the PIMC data.
}
\end{figure}

The corresponding density profile in coordinate space is depicted in Fig.~\ref{fig:density_qx2_A0p7}, with the green crosses again being the exact PIMC data. Evidently, the external perturbation dominates the behaviour of the system under these conditions and we find an almost shell-like structure with two pronounced peaks at the positions of the minima of the harmonic potential exceeding $2.5$ times the original unperturbed density $n_0$. These \emph{shells} are separated by two deep minima where the cosinusoidal potential has its maxima.

Let us next consider the solid red line showing the prediction by LRT, i.e., Eq.~(\ref{eq:coordinate_response_LRT}). Evidently, the deviation to the PIMC data is dramatic and almost attains $80\%$ in terms of the unperturbed density $n_0$. In particular, LRT predicts an unphysical negative density around the minima of the density profile. Furthermore, we find that the deviation profile (bottom panel) is quickly oscillating and exhibits a complicated behaviour, which means that the systematic deficiencies of LRT are both quantitative and qualitative.

The next step towards an improved description of $n(x)$ is given by the inclusion of the cubic response function $\chi^{(1,\textnormal{cubic})}(q)$ into $\braket{\hat\rho_\mathbf{q}}_{q,A}$, i.e., $\braket{\hat\rho_\mathbf{q}}_{q,A}\approx \chi^{(1)}(q)A + \chi^{(1,\textnormal{cubic})}(q)A^3$. The resulting evaluation of Eq.~(\ref{eq:expansion}) truncated at $\eta=1$ is depicted as the dash-double-dotted yellow curve and leads to a qualitative improvement of the description of the exact PIMC data, even though the deviations are still systematic and partly attain $\sim60\%$ of the unperturbed density $n_0$. 
This is somewhat expected as we know from Fig.~\ref{fig:PRL_rs2_theta1} that even the first harmonic is not properly described by $\chi^{(1)}(q)$ and $\chi^{(1,\textnormal{cubic})}(q)$ alone, and more terms in the expansion would be needed.

This is remedied by the dash-dotted black curve, where we again truncate Eq.~(\ref{eq:expansion}) after $\eta=1$, but include the full PIMC expectation value for $\braket{\hat\rho_\mathbf{q}}_{q,A}$. On the one hand, this does not substantially improve the agreement with the exact PIMC data and there remain systematic deviation of $\sim40\%$. On the other hand, it makes the deviation profile shown in the bottom panel more uniform. In particular, it exhibits oscillations with twice the wave number $q$ and a nearly constant amplitude.

Naturally, these deviations are caused by the second harmonic, and including the $\eta=2$ term with the exact PIMC expectation value for $\braket{\hat\rho_{2\mathbf{q}}}_{q,A}$ in Eq.~(\ref{eq:expansion}) leads to the dotted purple curve.  
The inclusion of the second harmonic leads to a drastic improvement, and the purple curve is in qualitative agreement with the green crosses everywhere. In particular, we note that the unphysical negative density points no longer appear. Still, the examination of the deviation profile of this curve shown in the bottom panel reveals that there remain systematic errors exceeding $5\%$. Further, the deviation has a rapidly oscillating form with a constant amplitude, and is readily identified as the contribution of the third harmonic, $k=3q$.

In other words, it is necessary to evaluate Eq.~(\ref{eq:expansion}) up to $\eta=3$ for such a large perturbation amplitude, and the results are shown as the dashed blue curve. Evidently, it is in excellent agreement with the PIMC data over the entire $x$-range, and the deviation plot shows that the difference between the two is fluctuating around zero within the given statistical uncertainty.

Let us next consider the right panel of Fig.~\ref{fig:spectrum_rs2}, where we show the $k$-dependence of the density response $\braket{\hat\rho_\mathbf{k}}_{q,A}$ for a smaller wave number, $q\approx0.84q_\textnormal{F}$. In particular, this is the smallest wave number that is accessible within a PIMC simulation for $N=14$, see also Sec.~\ref{sec:simulation_idea} above.

For the lowest depicted value of the perturbation amplitude ($A=0.05$, black squares), the signal at the first harmonic $\mathbf{k}=\mathbf{q}$ cannot be distinguished from LRT~\cite{Dornheim_PRL_2020}. In addition, we find a small yet significant signal for $\mathbf{k}=2\mathbf{q}$, which indicates that the second harmonic is the dominant contribution to nonlinear effects in the density response of electrons in the WDM regime. This is an important finding that is discussed extensively in the context of Fig.~\ref{fig:relative_rs2} below.

For $A=0.15$ (red circles), LRT still relatively accurately describes the signal at the first harmonic, and we find a deviation to the PIMC data of less than $1\%$. At the same time, the signal at $\mathbf{k}=2\mathbf{q}$ is significantly increased, which further corroborates the previous observation about the respective importance of the different harmonics to the nonlinear density response.

Upon further increasing the perturbation strength to $A=0.3$ (green crosses), the systematic error of LRT in the description of $\braket{\hat\rho_\mathbf{q}}_{q,A}$ increases to $\sim3\%$, while the second harmonic continues to have a higher impact. Moreover, we observe a small but significant density response at the third harmonic, that was absent for $A\lesssim0.15$.

The blue diamonds show $\braket{\hat\rho_\mathbf{k}}_{q,A}$ for $A=0.7$, which was the largest perturbation strength considered for $q\approx1.69q_\textnormal{F}$ shown in the left panel of Fig.~\ref{fig:spectrum_rs2}. For the smaller $q$, we observe a large systematic error of LRT for the first harmonic, which, however, is very accurately described by the cubic response function $\chi^{(1,\textnormal{cubic})}(q)$. This is a direct consequence of screening effects, which make nonlinear effects less important for small wave numbers, see Ref.~\cite{Dornheim_PRL_2020} for the first description of this finding. In the present work, we find similar effects for the second and third harmonics as well, see below.
Further, we find a large signal for $\mathbf{k}=2\mathbf{q}$ and a smaller, but noticeable signal for $\mathbf{k}=3\mathbf{q}$.

Finally, we consider the case of strong perturbation strength ($A=1.5$), which is depicted by the yellow triangles. At these parameters, the density response at the first harmonic is substantially reduced compared to LRT, and the signal at $\mathbf{k}=2\mathbf{q}$ is of the same order of magnitude. Further, the response at the third harmonic is drastically increased compared to $A=0.7$, and we even find a significant response for the fourth harmonic, $\mathbf{k}=4\mathbf{q}$.

\begin{figure}\centering
\includegraphics[width=0.485\textwidth]{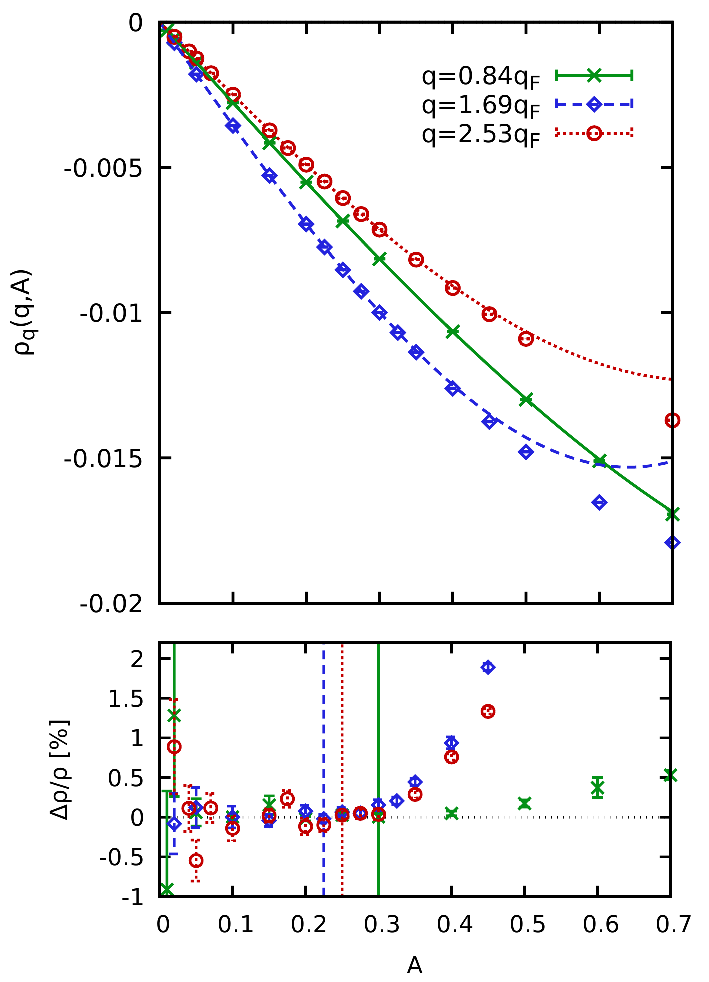}
\caption{\label{fig:determine_cubic} Determination of the cubic response function $\chi^{(1,\textnormal{cubic})}(q)$.
Top panel: density response of the UEG for $N=14$, $r_s=2$, and $\theta=1$ for $q\approx0.84q_\textnormal{F}$ (green crosses) $q\approx1.69q_\textnormal{F}$ (blue diamonds), and $q\approx2.53q_\textnormal{F}$ (red circles). The corresponding curves of the same colour have been obtained by fitting Eq.~(\ref{eq:fit}) to the PIMC data. Bottom panel: relative deviation of the fits to the PIMC data. The vertical lines show the respective maximum value of $A$ that was included in the fits.
}
\end{figure}

Let us proceed with a more rigorous quantification of the nonlinear electronic density response by obtaining the respective generalized response functions defined in Eqs.~(\ref{eq:fit})-(\ref{eq:third_fit}).
For the case of the cubic response function of the first harmonic $\chi^{(1,\textnormal{cubic})}(q)$, this is demonstrated in Fig.~\ref{fig:determine_cubic}
for three different wave numbers. More specifically, we show PIMC results for the dependence of $\braket{\hat\rho_\mathbf{q}}_{q,A}$ on the perturbation strength for $q\approx0.84q_\textnormal{F}$ (green crosses), $q\approx1.69q_\textnormal{F}$ (blue diamonds), and $q\approx2.53q_\textnormal{F}$ (red circles), and the corresponding curves show the respective fits according to Eq.~(\ref{eq:fit}).
The bottom panel shows the deviation between fits and PIMC data and the vertical lines indicate the respective maximum values of $A$ up to which data have been included in the fitting procedure.

Let us briefly touch upon the $A$-dependence of the Monte-Carlo error bars in the deviation plot.
For weak perturbations $A$, the absolute values of the density response $\braket{\hat\rho_\mathbf{q}}_{q,A}$ are small, which means that the denominator of the relative deviations shown in the bottom of Fig.~\ref{fig:determine_cubic} is small, too. At the same time, the absolute values of the Monte-Carlo error bars are nearly independent of $A$, which means that the relative accuracy of our PIMC data for Eq.~(\ref{eq:rho}) decreases for weak perturbations, and, eventually, the density response cannot be resolved within the given statistical uncertainty.

\begin{figure}\centering
\includegraphics[width=0.485\textwidth]{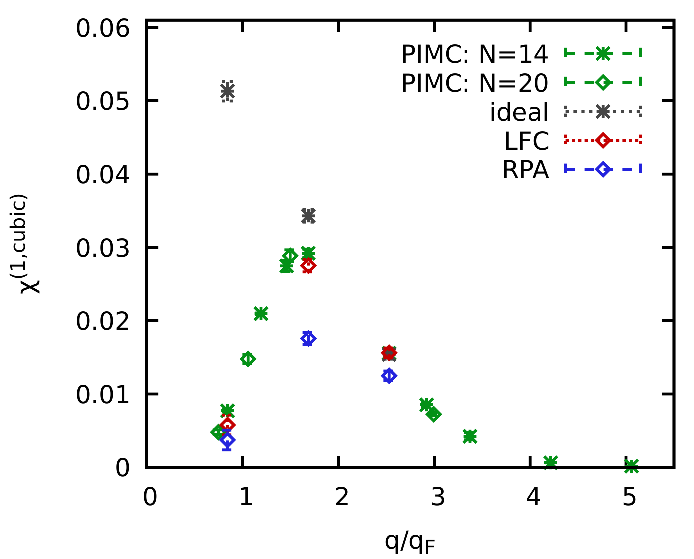}
\caption{\label{fig:cubic} Wave-number dependence of the cubic response function of the first harmonic $\chi^{(1,\textnormal{cubic})}(q)$ for the warm dense electron gas at $r_s=2$ and $\theta=1$.
}
\end{figure}

Regarding the $A$-dependence of $\braket{\hat\rho_\mathbf{q}}_{q,A}$ itself, we find that the blue diamonds exhibit the largest density response at the first harmonic, whereas the signal is weakest for the largest wave number, $q=2.53q_\textnormal{F}$. While not being trivial, this is somewhat expected as the LRT function $\chi^{(1)}(q)$, too, exhibits a maximum response around $q\sim1.5q_\textnormal{F}$, cf.~Fig.~\ref{fig:LRT_temperature} below. Interestingly, Eq.~(\ref{eq:fit}) remains accurate for larger perturbation amplitudes $A$ for the green crosses compared to the other two cases. This can be explained by the comparatively smaller value of $\chi^{(1,\textnormal{cubic})}$ in this case, see Fig.~\ref{fig:cubic}.

Repeating the fitting procedure demonstrated in Fig.~\ref{fig:determine_cubic} for different values of the perturbation wave numbers gives us access to the full $q$-dependence of $\chi^{(1,\textnormal{cubic})}(q)$, and the results are shown in Fig.~\ref{fig:cubic}. More specifically, the green data points have been obtained from our PIMC simulations and the stars and diamonds show results for $N=14$ and $N=20$ particles. While being available at different discrete $q$-points~\cite{dornheim_prl,dornheim_cpp17}, we note that no dependence on the system size can be resolved within the given confidence intervals~\cite{Dornheim_PRL_2020}. In addition, the grey stars, too, have been obtained from PIMC data, but for $N=14$ ideal (noninteracting) fermions. In this case, there are no screening effects and $\chi^{(1,\textnormal{cubic})}_0(q)$ attains a finite value in the limit of $q\to0$.

Unfortunately, no simple recursion relation is available for the ideal function $\chi_0^{(1,\textnormal{cubic})}(q)$ [see Sec.~\ref{sec:theory} above], so that we cannot evaluate the RPA and LFC expressions for the cubic response at the first harmonic given in Eqs.~(\ref{eq:cubic_first_RPA}) and (\ref{eq:cubic_first_LFC}) for all wave numbers. Yet, we can still use the three PIMC data points, and the results are depicted by the blue (RPA) and red (LFC) data points in Fig.~\ref{fig:cubic}. Evidently, the RPA curve seems to qualitatively reproduce the correct behaviour, but severely underestimates the true magnitude of the cubic response around its maximum. In contrast, the LFC curve nicely agrees with the PIMC data points for the interacting systems for all three wave numbers.

\begin{figure}\centering
\includegraphics[width=0.485\textwidth]{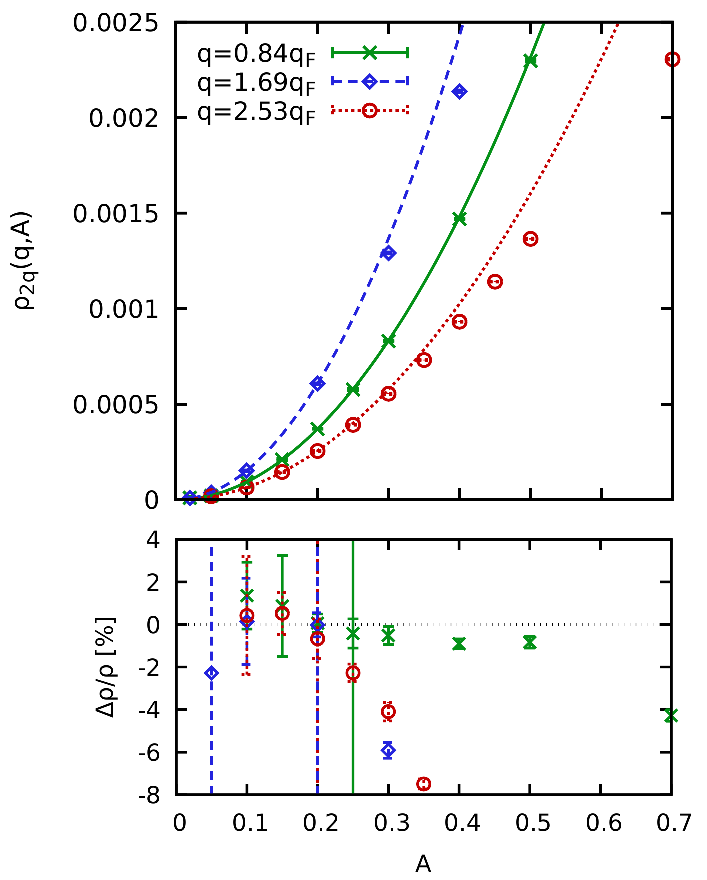}
\caption{\label{fig:determine_second} Determination of the quadratic response function $\chi^{(2)}(q)$.
Top panel: density response of the second harmonic $\braket{\hat\rho_{2\mathbf{q}}}_{q,A}$ of the UEG for $N=14$, $r_s=2$, and $\theta=1$ for $q\approx0.84q_\textnormal{F}$ (green crosses) $q\approx1.69q_\textnormal{F}$ (blue diamonds), and $q\approx2.53q_\textnormal{F}$ (red circles). The corresponding curves of the same colour have been obtained by fitting Eq.~(\ref{eq:second_fit}) to the PIMC data. Bottom panel: relative deviation of the fits to the PIMC data. The vertical lines show the respective maximum value of $A$ that was included in the fits. 
}
\end{figure}

Let us next repeat this analysis for the quadratic response function of the second harmonic $\chi^{(2)}(q)$, that has been shown to constitute the dominant nonlinear effect above. To this end, we show the $A$-dependence of the density response of the second harmonic $\braket{\hat\rho_{2\mathbf{q}}}_{q,A}$ in Fig.~\ref{fig:determine_second} for the same wave-numbers as in Fig.~\ref{fig:determine_cubic}.

Firstly, we observe the same absolute ordering of the three data sets with $q$ as for the first harmonic, with $q=1.69q_\textnormal{F}$ ($q=0.84q_\textnormal{F}$) exhibiting the largest (weakest) signal for the density response. Further, we see that the parabolic expansion given in Eq.~(\ref{eq:second_fit}) is in excellent agreement with the PIMC data points for small $q$, which gives us direct access to the corresponding quadratic density response function.

\begin{figure}\centering
\includegraphics[width=0.485\textwidth]{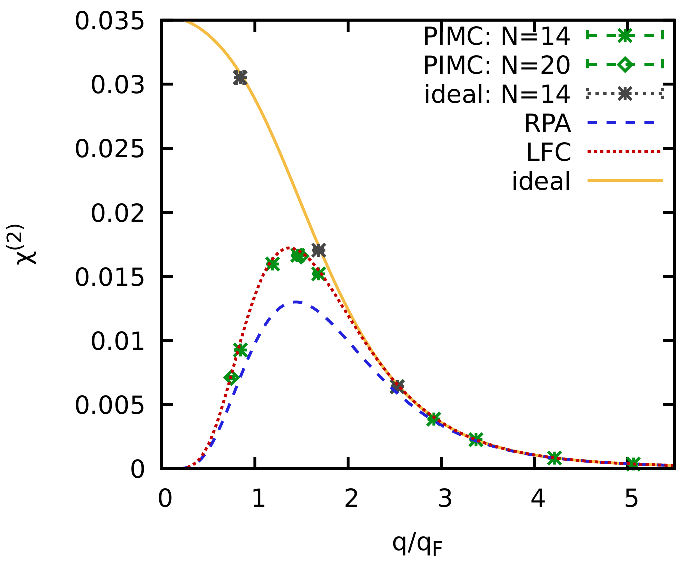}
\caption{\label{fig:second} Wave-number dependence of the quadratic response function of the second harmonic $\chi^{(2)}(q)$ for the warm dense electron gas at $r_s=2$ and $\theta=1$. The green stars (green diamonds) show PIMC results obtained from Eq.~(\ref{eq:second_fit}) for $N=14$ ($N=20$) electrons, and the grey stars to PIMC data for ideal (noninteracting) fermions. The dashed blue and dotted red lines show theoretical results within RPA [Eq.~(\ref{eq:chi2_RPA})] and using the static LFC from Ref.~\cite{dornheim_ML} [Eq.~(\ref{eq:chi2_LFC})], and the solid yellow line shows the ideal response function $\chi_0^{(2)}(q)$ [Eq.~(\ref{eq:Mikhailov2})].
}
\end{figure}

The results for the full wave-number dependence of $\chi^{(2)}(q)$ are shown in Fig.~\ref{fig:second} with the same key as $\chi^{(1,\textnormal{cubic})}(q)$ shown in Fig.~\ref{fig:cubic} above.
There is excellent agreement between the exact quadratic response function of the ideal Fermi gas [yellow curve, cf.~Eq.~(\ref{eq:Mikhailov2})] and our PIMC data for $N=14$ noninteracting fermions (grey stars) over the entire depicted $q$-range.
In addition, we find that the quadratic response function of the UEG $\chi^{(2)}(q)$ exhibits a qualitatively similar behaviour to $\chi^{(1,\textnormal{cubic})}(q)$ shown above. More specifically, $\chi^{(2)}(q)$ vanishes for small wave-numbers due to the perfect screening in the UEG, and, too, vanishes in the limit of large $q$, albeit slower than for $q\to0$.

The dashed blue RPA curve [Eq.~(\ref{eq:chi2_RPA})] is in qualitative agreement with our PIMC data for the UEG (green stars and diamonds) and correctly reproduces both of these limits. Yet, there appear systematic deviations exceeding $20\%$ for intermediate wave-numbers $q\sim1.5q_\textnormal{F}$ (i.e., in the vicinity of the maximum), which resembles the known deficiencies of RPA within LRT, see, e.g., Refs~\cite{dornheim_ML,dornheim_HEDP,dornheim_physrep_18}.

Finally, the dotted red curve has been obtained by including the neural-net representation of the static LFC $G(q)$ given by Dornheim \textnormal{et al.}~\cite{dornheim_ML} [Eq.~(\ref{eq:chi2_LFC})] and is in excellent agreement with the PIMC data for all $q$.

\begin{figure}\centering
\includegraphics[width=0.485\textwidth]{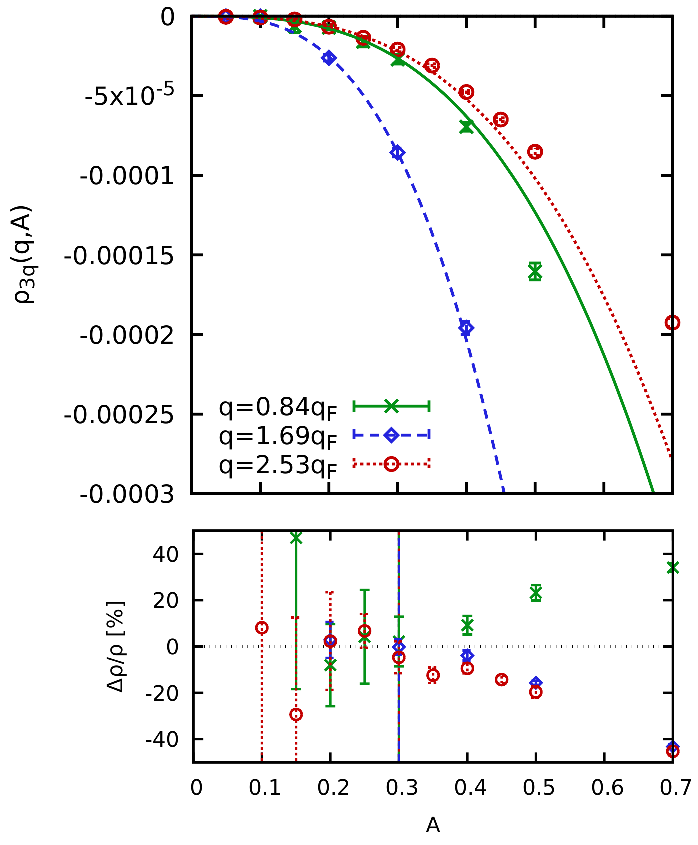}
\caption{\label{fig:determine_third} Determination of the cubic response function $\chi^{(3)}(q)$.
Top panel: density response of the third harmonic $\braket{\hat\rho_{3\mathbf{q}}}_{q,A}$ of the UEG for $N=14$, $r_s=2$, and $\theta=1$ for $q\approx0.84q_\textnormal{F}$ (green crosses) $q\approx1.69q_\textnormal{F}$ (blue diamonds), and $q\approx2.53q_\textnormal{F}$ (red circles). The corresponding curves of the same colour have been obtained by fitting Eq.~(\ref{eq:third_fit}) to the PIMC data. Bottom panel: relative deviation of the fits to the PIMC data. The vertical lines show the respective maximum value of $A$ that was included in the fits. 
}
\end{figure}

The last generalized response function to be considered in this work is the cubic response function of the third harmonic $\chi^{(3)}(q)$, that can be obtained by fitting Eq.~(\ref{eq:third_fit}) to  PIMC data for $\braket{\hat\rho_{3\mathbf{q}}}_{q,A}$. This is shown in Fig.~\ref{fig:determine_third}, where the $A$-dependence of the density response at $\mathbf{k}=3\mathbf{q}$ is shown for the same three wave numbers as for the other response functions investigated above. In this case, we observe a comparatively large increase of $\braket{\hat\rho_\mathbf{3q}}_{q,A}$ with $A$ at $q=1.69q_\textnormal{F}$ that is approximately four times as large as for the other two wave-numbers, which are similar in magnitude.

Moreover, we note that the absolute value of the density response at $\mathbf{k}=3\mathbf{q}$ is an order of magnitude smaller than for $\mathbf{k}=2\mathbf{q}$ for the same values of the perturbation amplitude $A$. This, in turn, means that the relative error bars in $\braket{\hat\rho_\mathbf{3q}}_{q,A}$ are larger than for the other harmonics shown above, which makes the determination of $\chi^{(3)}(q)$ more challenging.

\begin{figure}\centering
\includegraphics[width=0.485\textwidth]{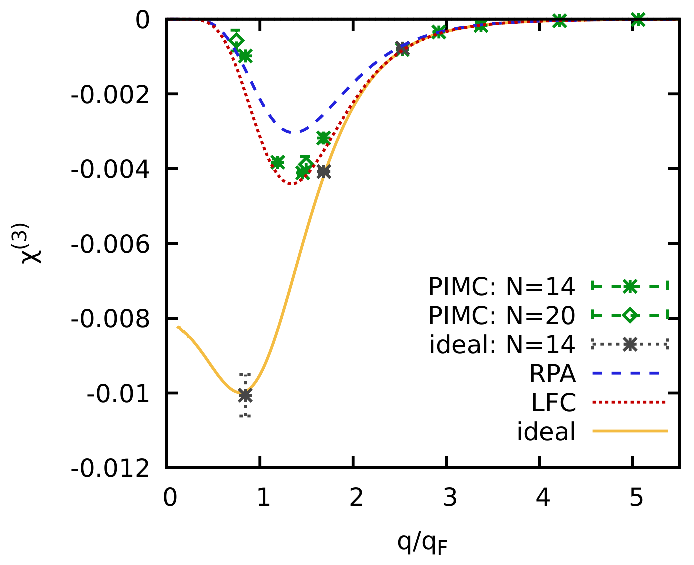}
\caption{\label{fig:third} Wave-number dependence of the cubic response function of the third harmonic $\chi^{(3)}(q)$ for the warm dense electron gas at $r_s=2$ and $\theta=1$. The green stars (green diamonds) show PIMC results obtained from Eq.~(\ref{eq:third_fit}) for $N=14$ ($N=20$) electrons, and the grey stars to PIMC data for ideal (noninteracting) fermions. The dashed blue and dotted red lines show theoretical results within RPA [Eq.~(\ref{eq:chi3_RPA})] and using the static LFC from Ref.~\cite{dornheim_ML} [Eq.~(\ref{eq:chi3_LFC})], and the solid yellow line shows the ideal response function $\chi_0^{(3)}(q)$ [Eq.~(\ref{eq:Mikhailov3})].}
\end{figure}

The corresponding wave-number dependence of this function is shown in Fig.~\ref{fig:third}. Let us first focus on the noninteracting case, with the grey stars depicting PIMC results for $N=14$ ideal fermions and the solid yellow curve the exact ideal response function $\chi_0^{(3)}(q)$ that we obtain from Eq.~(\ref{eq:Mikhailov3}). Remarkably, even without the electronic Coulomb repulsion and the associated screening effects, the density response function of the third harmonic exhibits a complicated, non-monotonous behaviour, with a maximum in magnitude around the Fermi wave number. This is confirmed by the corresponding ideal PIMC data point at $q=0.84q_\textnormal{F}$, which nicely agrees with the exact result.

For the interacting UEG, this interesting behaviour at small $q$ is masked by screening effects, and the generalized response function exhibits similar trends as $\chi^{(1,\textnormal{cubic})}(q)$ and $\chi^{(2)}(q)$, that is, it vanishes both in the limits of small and large $q$. As seen before, the RPA of the cubic response at the third harmonic underestimates the response, but the inclusion of LFCs gives a nice agreement between QMC simulations and theory.

\begin{figure}\centering
\includegraphics[width=0.485\textwidth]{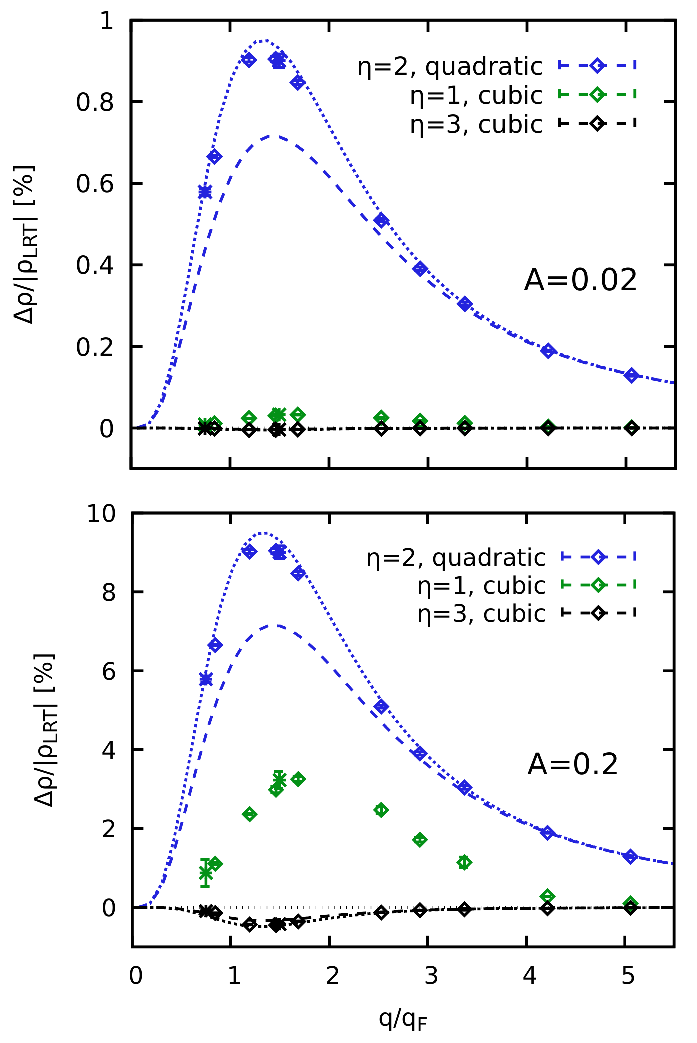}
\caption{\label{fig:relative_rs2}
Relative impact of different nonlinear contributions to the density response $\braket{\hat\rho_\mathbf{q}}$ compared to the prediction for $\braket{\hat\rho_\mathbf{q}}_{q,A}$ from LRT for $A=0.02$ (top) and $A=0.2$ (bottom). Shown are $\braket{\hat\rho_{2\mathbf{q}}}_{q,A}=\chi^{(2)}(q)A^2$ (blue), $\braket{\hat\rho_{3\mathbf{q}}}_{q,A}=\chi^{(3)}(q)A^3$ (black), and $\braket{\Delta\hat\rho_\mathbf{q}}_{q,A}=\chi^{(1,\textnormal{cubic})}(q)A^3$ (green). The dashed and dotted lines have been obtained within RPA and using a LFC, respectively. 
}
\end{figure}

Let us conclude this section with a systematic investigation of the respective contribution of the different generalized response functions to the total nonlinear density response. In Fig.~\ref{fig:relative_rs2}, we show the relative contribution of $\chi^{(2)}(q)$ (blue), $\chi^{(1,\textnormal{cubic})}(q)$ (green), and $\chi^{(3)}(q)$ (black) compared to the density response at the first harmonic predicted by LRT, Eq.~(\ref{eq:LRT}). More specifically, the data points have been obtained from fits to PIMC data, and the dotted (dashed) lines are theoretical results including an LFC (RPA only).

The top panel of Fig.~\ref{fig:relative_rs2} corresponds to a weak perturbation, $A=0.02$. At these conditions, LRT is relatively accurate (see, e.g., the left panel of Fig.~\ref{fig:density_qx2}) and the response due to the cubic response functions for both the first and third harmonic is smaller than $0.1\%$ compared to LRT. By far the largest nonlinear effect can be observed at the second harmonic with a maximum signal of almost $1\%$ in terms of $\rho_\textnormal{LRT}$. In particular, the response for $\mathbf{k}=2\mathbf{q}$ is more than an order of magnitude larger than the other two depicted curves.

The bottom panel of Fig.~\ref{fig:relative_rs2} corresponds to a stronger perturbation, $A=0.2$. This is a particularly interesting example, as the density response for the first three harmonics is fully described by the different response functions at these conditions. In contrast, additional terms of the respective expansions in powers of $A$ would be needed for an accurate description of $\braket{\hat\rho_\mathbf{k}}_{q,A}$ at much larger values of $A$.

Let us first consider the density response of the second harmonic which, again, constitutes the largest nonlinear effect, with a maximum contribution of almost $10\%$ around $q\sim1.5q_\textnormal{F}$ compared to Eq.~(\ref{eq:LRT}). Further, we note that the relative impact of $\braket{\hat\rho_{2\mathbf{q}}}_{q,A}$ only diminishes in the small wave-number limit for $q\lesssim0.2q_\textnormal{F}$, and disappears even more slowly for large values of $q$, with $\braket{\hat\rho_{2\mathbf{q}}}_{q,A}/\rho_\textnormal{LRT}$ exceeding $1\%$ even at $q=5q_\textnormal{F}$.

In addition, the cubic contribution to the density response at the first harmonic $\braket{\Delta\hat\rho_\mathbf{q}}_{q,A}=\chi^{(1,\textnormal{cubic})}(q)A^3$ is substantially more important compared to $A=0.02$, and we find a maximum impact of approximately $4\%$ around $q=1.5q_\textnormal{F}$. Indeed, this is the same order of magnitude as the signal of the second harmonic, which means that both $\chi^{(2)}(q)$ and $\chi^{(1,\textnormal{cubic})}(q)$ have to be taken into account for an accurate description of the system, cf.~the right panel of Fig.~\ref{fig:density_qx2} above showing the radial density $n(\mathbf{r})$ for this case. At the same time, we find that the impact of the cubic contribution to the first harmonic rapidly vanishes in both the limits of large and small wave-numbers and is only relevant for intermediate $q$.

Finally, the black points and curves correspond to the contribution of the third harmonic, which is an order of magnitude smaller than the green curve. Therefore, it appears to be mostly negligible for practical purposes.

\subsection{Temperature dependence\label{sec:temperature}}

\begin{figure}\centering
\includegraphics[width=0.475\textwidth]{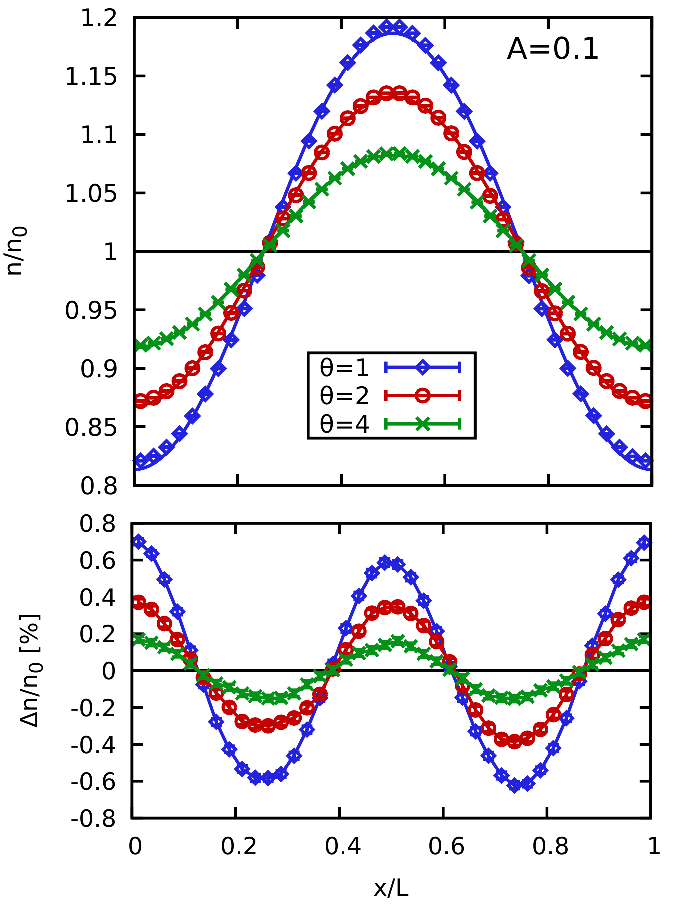}
\caption{\label{fig:temperature_density_A0p1}
Density profile of the harmonically perturbed electron gas for $N=14$ and $r_s=2$, with $q\approx0.84q_\textnormal{F}$ and $\mathbf{q}=2\pi/L(1,0,0)^T$ for $A=0.1$. Top: PIMC results for the density $n(x)$ along the direction of the perturbation for $\theta=1$ (blue diamonds), $\theta=2$ (red circles), and $\theta=4$ (green crosses). The solid curves correspond to LRT [cf.~Eq.~(\ref{eq:coordinate_response_LRT})].
Bottom: relative deviation of LRT to the PIMC data.
}
\end{figure}

Let us next turn our attention to the dependence of nonlinear effects in the static density response on the temperature $\theta$. In Fig.~\ref{fig:temperature_density_A0p1}, we show the density profile in coordinate space, $n(\mathbf{r})$, along the direction of the perturbation, for $N=14$ electrons at $r_s=2$ for $q\approx0.84q_\textnormal{F}$ and a relatively small value of the perturbation amplitude, $A=0.1$. The blue diamonds have been obtained for $\theta=1$ and exhibit a deviation from the unperturbed density $n_0$ of almost $20\%$ at the maxima and minima. The solid blue curve shows the prediction from LRT [cf.~Eq.~(\ref{eq:coordinate_response_LRT})] and qualitatively captures the correct behaviour. This can again be seen particularly well in the bottom panel where we show the relative deviation between the PIMC data and LRT. Evidently, LRT results in an oscillating deviation profile that is mainly due to the density response at the second harmonic $\braket{\hat\rho_{2\mathbf{q}}}_{q,A}$ (see the previous section \ref{sec:metallic}), with a maximum deviation below $0.8\%$.

Let us next focus on the red circles corresponding to twice the temperature, $\theta=2$. Firstly, we note that the overall density response is reduced compared to $\theta=1$, as it is expected, see also the corresponding discussion in Refs.~\cite{Dornheim_PRL_2020,dornheim_HEDP}. In particular, an increasing temperature leads to an increased internal energy scale of the system, which, in turn, means that the relative impact of an external potential is reduced if the perturbation amplitude $A$ is kept constant. The green crosses correspond to a further increase of the temperature by a factor of two, and the total density response is again reduced compared to the other data sets.

In addition, a reduced density response means that nonlinear effects become less important and we find maximum deviations between LRT and the PIMC data of $0.4\%$ ($0.2\%$) for $\theta=2$ ($\theta=4$). At the same time, we note that the nonlinear response is significant for all three depicted temperatures and is mainly given by the signal at the second harmonic, $\mathbf{k}=2\mathbf{q}$.

\begin{figure}\centering
\includegraphics[width=0.475\textwidth]{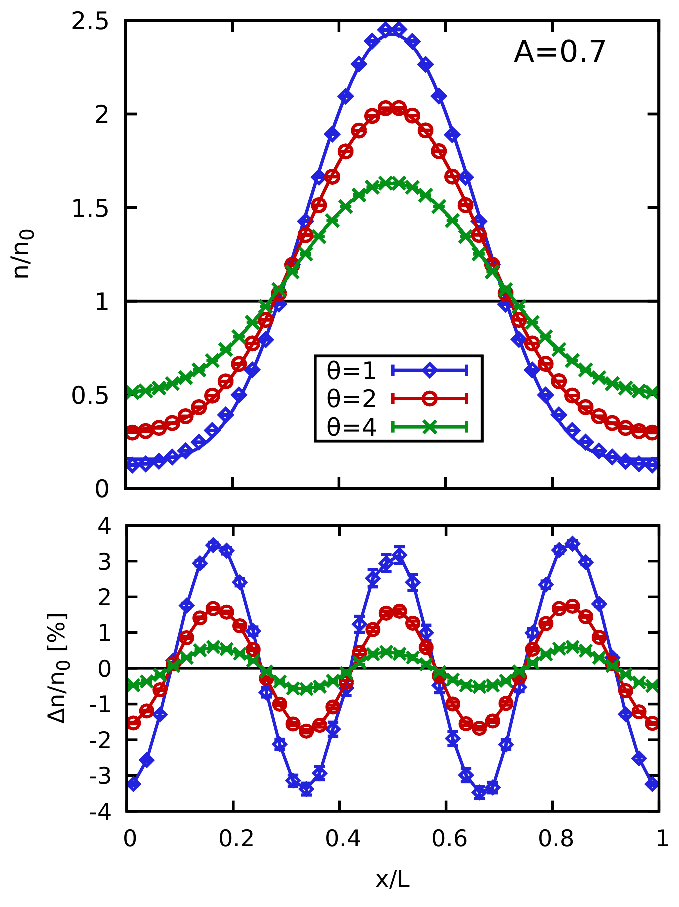}
\caption{\label{fig:temperature_density_A0p7}
Density profile of the harmonically perturbed electron gas for $N=14$ and $r_s=2$, with $q\approx0.84q_\textnormal{F}$ and $\mathbf{q}=2\pi/L(1,0,0)^T$ for $A=0.7$. Top: PIMC results for the density $n(x)$ along the direction of the perturbation for $\theta=1$ (blue diamonds), $\theta=2$ (red circles), and $\theta=4$ (green crosses). The solid curves correspond to the evaluation of Eq.~(\ref{eq:expansion}) truncated after $\eta=2$.
Bottom: relative deviation of solid curves from the PIMC data.
}
\end{figure}

Let us next consider Fig.~\ref{fig:temperature_density_A0p7}, which shows the same information as Fig.~\ref{fig:temperature_density_A0p1}, but for a significantly increased perturbation strength $A=0.7$. In this case, the density response is strongly nonlinear for all three depicted temperatures, although the density profile $n(\mathbf{r})$ is substantially different for all three cases. For $\theta=1$, the external perturbation constitutes the dominant effect and fully shapes the physical behaviour of the system. Most particles are, on average, located around the center of the simulation cell where the cosinusoidal potential has its minimum, whereas the density is reduced by roughly $90\%$ for small and large $x$. The solid blue curve has been computed by truncating Eq.~(\ref{eq:expansion}) after the second harmonic ($\eta=2$) using the exact PIMC expectation values for the coefficients $\braket{\hat\rho_\mathbf{q}}_{q,A}$ and $\braket{\hat\rho_{2\mathbf{q}}}_{q,A}$. In this case, we find that the agreement to the PIMC data is only very qualitative, and there appear rapidly oscillating deviations (bottom panel) with an amplitude exceeding $3\%$. Naturally, these deviations constitute the signal of the third harmonic, $\mathbf{k}=3\mathbf{q}$, see also the discussion of Fig.~\ref{fig:spectrum_temperature} below.

For $\theta=2$, the density at the maxima of the external perturbation is substantially larger compared to the lower temperature, and the impact of the third harmonic on the density profile $n(\mathbf{r})$ does not exceed $2\%$. 
Finally at the highest temperature, we find oscillations in the total density $n$ of approximately $50\%$, which is substantially smaller compared to the other cases. Consequently, the first two harmonics already provide a rather accurate description, and the systematic error is smaller than $1\%$ everywhere.

\begin{figure}\centering
\includegraphics[width=0.485\textwidth]{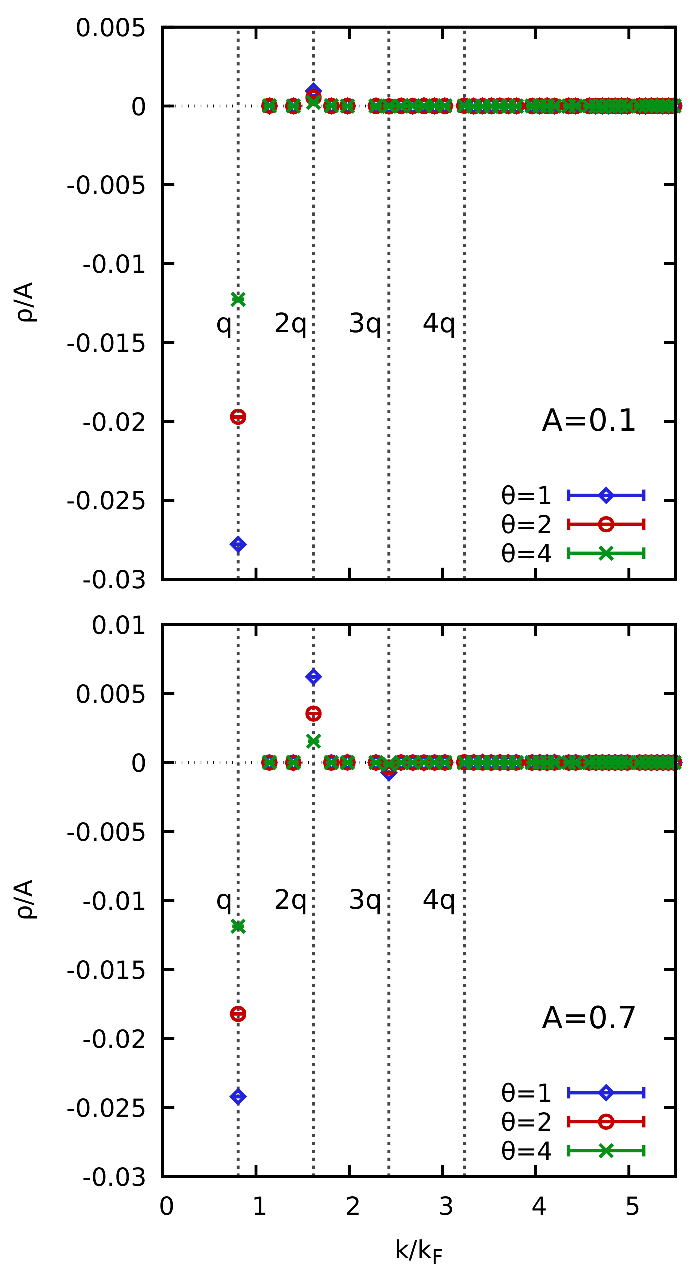}
\caption{\label{fig:spectrum_temperature}
Full wave-number dependence of the density response $\braket{\hat\rho_\mathbf{k}}_{q,A}$ of the UEG for $N=14$, $r_s=2$, and $\theta=1$ with $q\approx0.84q_\textnormal{F}$ and $A=0.1$ (top) and $A=0.7$ (bottom). Shown are PIMC results for Eq.~(\ref{eq:rho}) for $\theta=1$ (blue diamonds), $\theta=2$ (red circles), and $\theta=4$ (green crosses). The vertical dotted lines indicate the location of the first four harmonics.
}
\end{figure}

A more systematic investigation of the respective strength of the density response at different harmonics is shown in Fig.~\ref{fig:spectrum_temperature}, where we show the full $\mathbf{k}$-dependence of $\braket{\hat\rho_\mathbf{k}}_{q,A}$ for the two examples investigated in Figs.~\ref{fig:temperature_density_A0p1} and \ref{fig:temperature_density_A0p7}. The top panel corresponds to the weaker perturbation, $A=0.1$.
We note that we again only observe a nonzero signal for the integer harmonics of $\mathbf{q}$, as it is expected. For the first harmonic ($\mathbf{k}=\mathbf{q}$), we see the expected order of the signals for the three depicted temperatures, with the response at $\theta=4$ being less than half the response at $\theta=1$.

For the second harmonic ($\mathbf{k}=2\mathbf{q}$), the relative strength of the signals for different temperatures is even more different, with the response for $\theta=4$ ($\theta=2$) being approximately $25\%$ ($50\%$) of the response for $\theta=1$.

Finally, we mention that the signal at the third harmonic ($\mathbf{k}=3\mathbf{q}$) vanishes within the given level of accuracy for all $\theta$.

The bottom panel of Fig.~\ref{fig:spectrum_temperature} shows the same results for $A=0.7$ and we find similar trends. In fact, the ratio of density response for $\theta=2$ and $\theta=4$ for both the first and second harmonic compared to the respective response at $\theta=1$ is remarkably unchanged, although the total response at $\mathbf{k}=2\mathbf{q}$ has substantially increased. Moreover, we find a nonzero signal at $\mathbf{k}=3\mathbf{q}$ for all three temperatures, whereas the response at the fourth harmonic vanishes even for $\theta=1$.

\begin{figure}\centering
\includegraphics[width=0.485\textwidth]{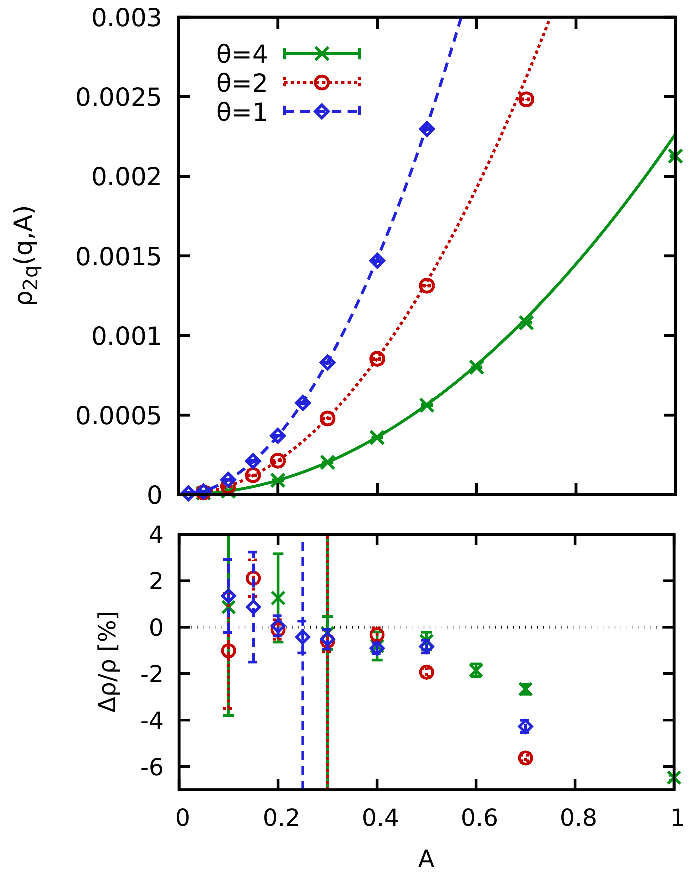}
\caption{\label{fig:temperature_second} Determination of the quadratic response function $\chi^{(2)}(q)$.
Top panel: density response of the second harmonic $\braket{\rho_{2\mathbf{q}}}_{q,A}$ of the UEG for $N=14$, $r_s=2$, and $q\approx0.84q_\textnormal{F}$ for $\theta=1$ (blue diamonds) $\theta=2$ (red circles), and $\theta=4$ (green crosses). The corresponding curves of the same colour have been obtained by fitting Eq.~(\ref{eq:second_fit}) to the PIMC data. Bottom panel: relative deviation of the fits to the PIMC data. The vertical lines show the respective maximum value of $A$ that was included in the fits. 
}
\end{figure}

The next step towards a more complete description of the temperature dependence of the nonlinear density response is given by the determination of the different generalized response functions. To this end, we show the dependence of the density response at the second harmonic $\braket{\hat\rho_{2\mathbf{q}}}_{q,A}$ on the perturbation amplitude $A$ in Fig.~\ref{fig:temperature_second} for the same conditions as in the previous figures in this section. The points depict PIMC data for the three different temperatures, and the curves parabolic fits to these data according to Eq.~(\ref{eq:second_fit}). The bottom panel shows the relative deviation between PIMC and the fits, and the vertical bars show the respective maximum value of $A$ up to which points have been included into the fitting procedure. Evidently, Eq.~(\ref{eq:second_fit}) is capable to accurately reproduce the PIMC data for $A\lesssim0.4$ in all three cases.
Going back to the top panel, we see that both the PIMC results and the parabolic fits are systematically ordered with $\theta$, as it is expected.

\begin{figure}\centering
\includegraphics[width=0.485\textwidth]{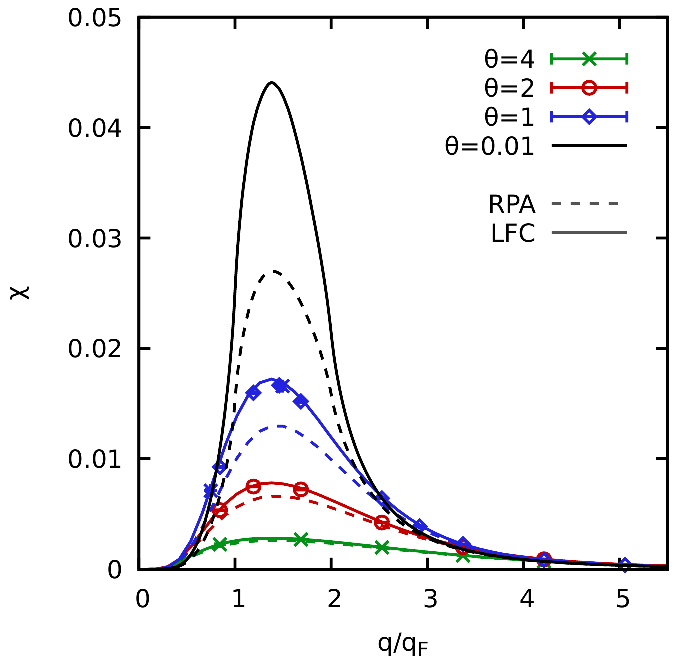}
\caption{\label{fig:quadratic_temperature} Wave-number dependence of the quadratic response function of the second harmonic $\chi^{(2)}(q)$ for the warm dense electron gas at $r_s=2$ and $\theta=1$ (blue diamonds), $\theta=2$ (red circles), and $\theta=4$ (green crosses).  The dashed and solid lines show theoretical results within RPA and using the LFC from Ref.~\cite{dornheim_ML}, cf.~Eqs.~(\ref{eq:chi2_RPA}) and (\ref{eq:quadratic_LFC}).
}
\end{figure}

Repeating this analysis for different values of the perturbation wave-number $q$ gives us access to the full wave-number dependence of the quadratic response function of the second harmonic $\chi^{(2)}(q)$, and the results are shown in Fig.~\ref{fig:quadratic_temperature} for different relevant temperatures. The symbols have been obtained on the basis of fitting Eq.~(\ref{eq:second_fit}) to PIMC data, and the solid [dashed] curves correspond to the theory with a LFC [within RPA], cf.~Eq.~(\ref{eq:chi2_LFC}) [Eq.~(\ref{eq:chi2_RPA})].

First and foremost, we note that $\chi^{(2)}$ exhibits a substantial dependence on $\theta$, and the maximum value of this function at $\theta=0.01$ (black) is larger by an order of magnitude compared to $\theta=4$ (green). Moreover, the shape of the curves strongly depends on $\theta$ as well: while the curves for all temperatures are relatively similar both in the limit of small and large wave-numbers $q$, there emerges a sharp peak around $q\sim1.3q_\textnormal{F}$ with decreasing $\theta$.

Let us next analyse the level of accuracy of both RPA and LFC, which, too, strongly depends on $\theta$. For $\theta=4$, the system is only weakly coupled and both RPA and LFC are in very good agreement to the PIMC data (green crosses) over the entire depicted wave-number range. Upon decreasing the temperature to $\theta=2$, RPA systematically underestimates the quadratic response function in the vicinity of its maximum, whereas the LFC still nicely reproduces the PIMC results (red circles) everywhere. For $\theta=1$, the deviation between RPA and LFC becomes substantial and partly exceeds $20\%$, whereas the LFC gives both a qualitatively and quantitatively sufficient description. At the same time, we like to re-iterate the point that Eq.~(\ref{eq:chi2_LFC}) does only include screening effects to a linear order, and, thus does not give an exact description even with the exact static limit of the LFC $G(q)$ taken from Ref.~\cite{dornheim_ML}. Indeed, the systematic inaccuracy of the approximate description of screening effects in Eq.~(\ref{eq:chi2_LFC}) increases with the effective coupling strength of the system. With the density parameter $r_s$ being constant, the effective coupling increases towards small temperature and small deviations between the PIMC data points and the theoretical prediction become noticeable for $\theta\lesssim1$.

\begin{figure}\centering
\includegraphics[width=0.485\textwidth]{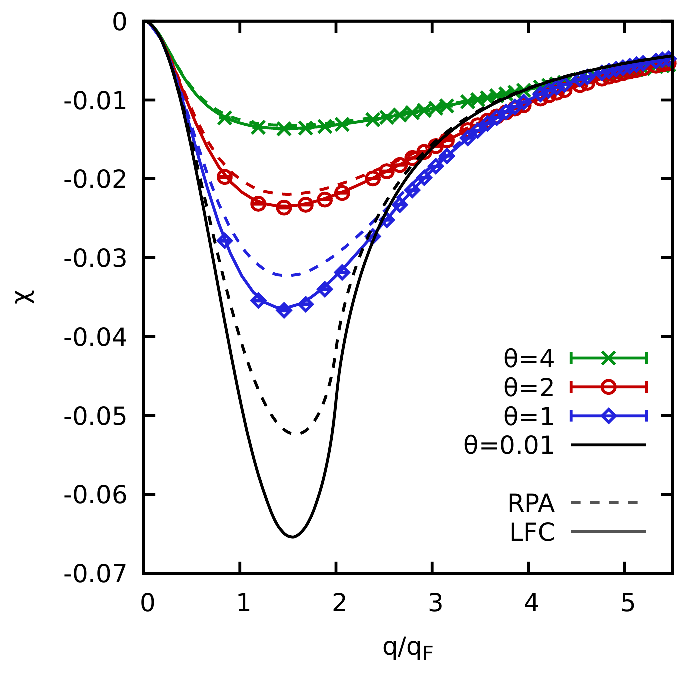}
\caption{\label{fig:LRT_temperature} Wave-number dependence of the linear response function $\chi^{(1)}(q)$ for the warm dense electron gas at $r_s=2$ and $\theta=1$ (blue diamonds), $\theta=2$ (red circles), and $\theta=4$ (green crosses). The dashed and solid lines have been obtained within RPA and using the LFC from Ref.~\cite{dornheim_ML}, see Eq.~(\ref{eq:LFC}).
}
\end{figure}

\begin{figure}\centering
\includegraphics[width=0.485\textwidth]{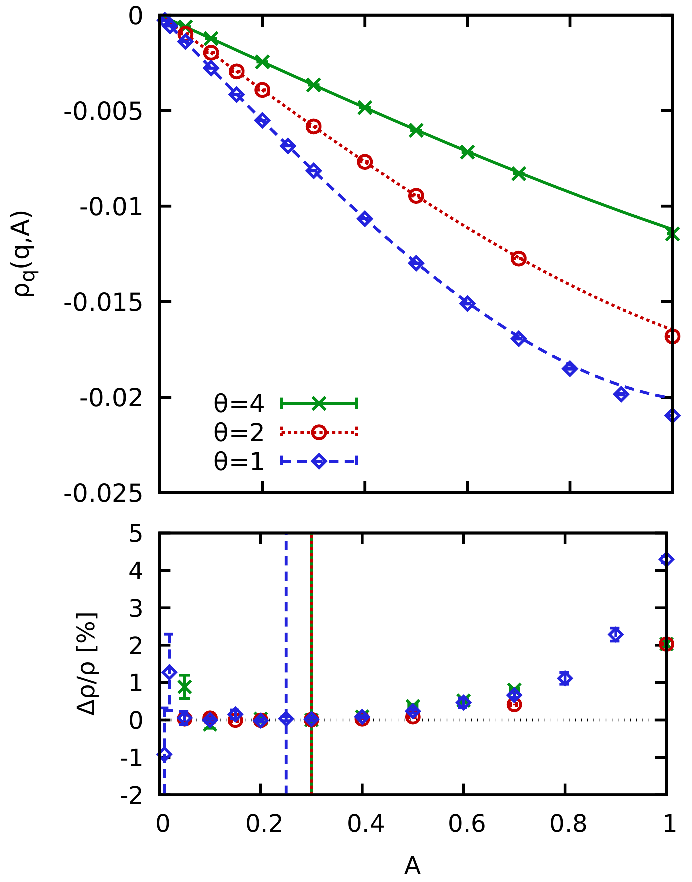}
\caption{\label{fig:temperature_first} Determination of the cubic response function $\chi^{(1,\textnormal{cubic})}(q)$.
Top panel: density response of the first harmonic $\braket{\rho_{\mathbf{q}}}_{q,A}$ of the UEG for $N=14$, $r_s=2$, and $q\approx0.84q_\textnormal{F}$ for $\theta=1$ (blue diamonds) $\theta=2$ (red circles), and $\theta=4$ (green crosses). The corresponding curves of the same color have been obtained by fitting Eq.~(\ref{eq:fit}) to the PIMC data. Bottom panel: relative deviation of the fits to the PIMC data. The vertical lines show the respective maximum value of $A$ that was included in the fits. 
}
\end{figure}

Finally, the black curves show results for $\theta=0.01$, where the system is in the ground state. In this case, the systematic deviation between LFC and RPA is even larger compared to $\theta=1$, as the effective coupling strength of the system is further increased. Unfortunately, PIMC simulations are not feasible at these conditions due to the fermion sign problem~\cite{dornheim_sign_problem}, and ground-state QMC simulations are not available for this quantity.

As a reference, we show the analogous temperature dependence of the LRT response function $\chi^{(1)}(q)$ in Fig.~\ref{fig:LRT_temperature}. Overall, $\chi^{(1)}(q)$ exhibits a qualitatively similar behaviour as $\chi^{(2)}(q)$, with a screening-induced decay at small wave-numbers, and a maximum of its modulus around $q\sim1.5q_\textnormal{F}$. Still, the decrease of the density response towards high temperature is substantially less pronounced for $\chi^{(1)}(q)$, which makes the experimental probing of the second harmonic a potentially valuable method of diagnostic for the temperature~\cite{siegfried_review,kraus_xrts}.

The next generalized response function of interest is given by $\chi^{(1,\textnormal{cubic})}(q)$, the determination of which is demonstrated in Fig.~\ref{fig:temperature_first} for the usual three values of $\theta$. The data points are PIMC results for $\braket{\hat\rho_\mathbf{q}}_{q,A}$ and the different curves have been fitted to the latter via Eq.~(\ref{eq:fit}). Evidently, this functional form is again well suited to represent the PIMC data, and deviations can only be spotted with the naked eye at large perturbation amplitudes and are most pronounced for $\theta=1$. In the bottom panel, we show the corresponding relative deviations $\Delta\rho/\rho$, with the vertical lines representing the maximum values of the perturbation amplitude $A$ that have been included into the fitting procedure. In particular, the fit function remains accurate for large values of $A$ beyond the respective fitting interval for all three temperatures, which is a strong indication for the high quality of our approach.

\begin{figure}\centering
\includegraphics[width=0.485\textwidth]{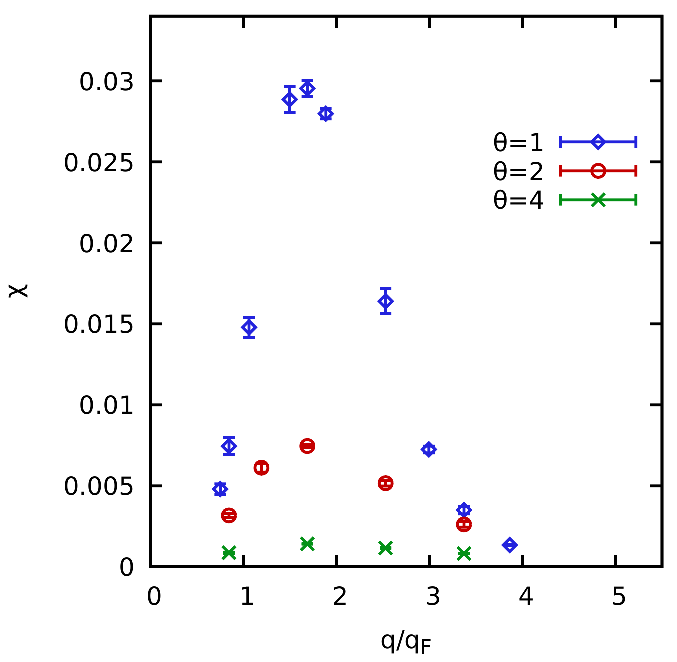}
\caption{\label{fig:cubic_temperature} Wave-number dependence of the cubic response function of the first harmonic $\chi^{(1,\textnormal{cubic})}(q)$ for the warm dense electron gas at $r_s=2$ and $\theta=1$ (blue diamonds), $\theta=2$ (red circles), and $\theta=4$ (green crosses).  
}
\end{figure}

The full wave-number dependence of the thus determined cubic response function of the first harmonic $\chi^{(1,\textnormal{cubic})}(q)$ is shown in Fig.~\ref{fig:cubic_temperature}.
First and foremost, we find that this function exhibits an even more pronounced dependence on the temperature compared to both $\chi^{(1)}(q)$ and $\chi^{(2)}(q)$, which further highlights the potential utility of the nonlinear density response as a method of diagnostics for WDM. More specifically, the magnitude of the maximum at $\theta=4$ is more than order of magnitude smaller compared to its analogue at $\theta=1$.
Unfortunately, no RPA and LFC data are available in this case, as the ideal cubic response function is not known analytically and, hence, would require extensive PIMC simulations of the harmonically perturbed ideal Fermi gas.

\begin{figure}\centering
\includegraphics[width=0.485\textwidth]{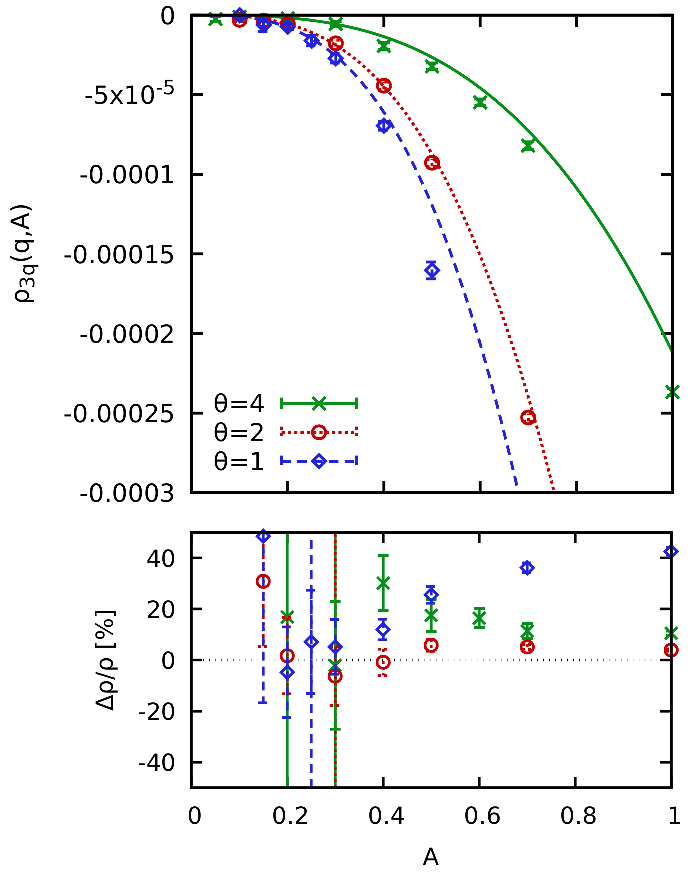}
\caption{\label{fig:temperature_third} Determination of the cubic response function $\chi^{(3)}(q)$.
Top panel: density response of the third harmonic $\braket{\rho_{3\mathbf{q}}}_{q,A}$ of the UEG for $N=14$, $r_s=2$, and $q\approx0.84q_\textnormal{F}$ for $\theta=1$ (blue diamonds) $\theta=2$ (red circles), and $\theta=4$ (green crosses). The corresponding curves of the same colour have been obtained by fitting Eq.~(\ref{eq:second_fit}) to the PIMC data. Bottom panel: relative deviation of the fits from the PIMC data. The vertical lines show the respective maximum value of $A$ that was included in the fits. 
}
\end{figure}

The final response function considered in this work is given by the cubic response function of the third harmonic, $\chi^{(3)}(q)$. The determination of this function on the basis of our PIMC data for $\braket{\hat\rho_{3\mathbf{q}}}_{q,A}$ is shown in Fig.~\ref{fig:temperature_third}, where the curves have been fitted to the latter via Eq.~(\ref{eq:third_fit}). Since the observed signal for the third harmonic is comparably quite small for all three temperatures (cf.~Fig.~\ref{fig:spectrum_temperature}), the relative statistical uncertainty of the PIMC data points is large, especially at small $A$. Yet, Eq.~(\ref{eq:third_fit}) only constitutes the lowest order in $A$ of $\braket{\hat\rho_{3\mathbf{q}}}_{q,A}$, which makes the determination of the coefficient $\chi^{(3)}(q)$ difficult.
This is particularly true for $\theta=4$, where we can only give a qualitative estimation of the order of magnitude of this function.

\begin{figure}\centering
\includegraphics[width=0.485\textwidth]{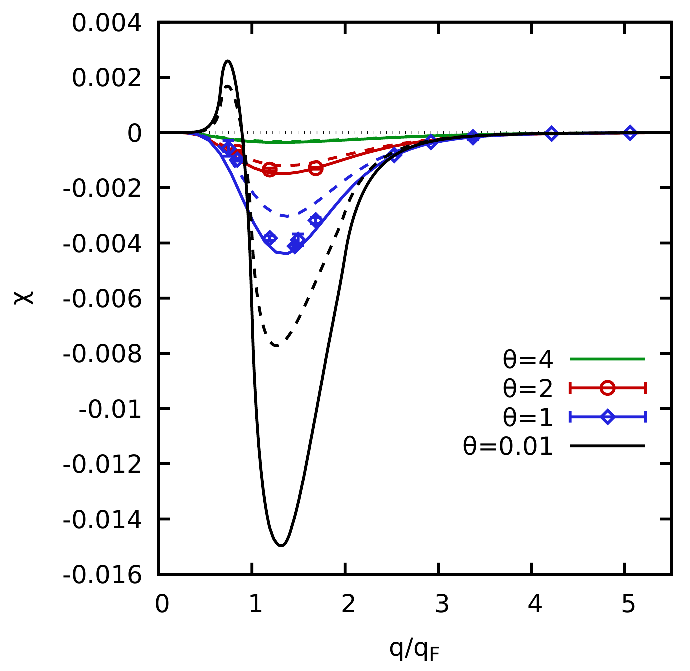}
\caption{\label{fig:third_temperature} Wave-number dependence of the cubic response function of the third harmonic $\chi^{(3)(q)}$ for the warm dense electron gas at $r_s=2$ and $\theta=1$ (blue diamonds), $\theta=2$ (red circles), and $\theta=4$ (green crosses). The solid and dashed lines have been obtained using the LFC from Ref.~\cite{dornheim_ML} and within RPA, respectively.
}
\end{figure}

The full wave-number dependence of $\chi^{(3)}(q)$ is shown in Fig.~\ref{fig:third_temperature}, again for a gamut of different values of $\theta$. For the highest temperature ($\theta=4$, green curve), the signal is two orders of magnitude smaller than in the ground-state ($\theta=0.01$, black curve) and can hardly be resolved with the bare eye. Upon decreasing the temperature to $\theta=2$ (red curve), the signal is increased by a factor of five and can be resolved with the PIMC procedure described in the discussion of Fig.~\ref{fig:temperature_third}. Moreover, the difference between RPA [Eq.~(\ref{eq:chi3_RPA})] and the LFC-based description [Eq.~(\ref{eq:chi3_LFC})] is quite small, and the PIMC data points are in excellent agreement with the latter.

Decreasing the temperature by an additional factor of two leads to the blue curves and data points, which display approximately thrice the magnitude in the maximum of the density response around $q\sim1.5q_\textnormal{F}$. In addition, there appears a much larger difference between RPA and LFC, and---although to a much smaller degree---between the LFC curve and the PIMC data; see also the discussion of Fig.~\ref{fig:cubic} above.

Lastly, the black curves correspond to the ground state ($\theta=0.01$), where no PIMC simulations are available due to the fermion sign problem~\cite{dornheim_sign_problem}. Evidently, RPA underestimates the depth of the minimum of $\chi^{(3)}(q)$ around $q\approx1.5q_\textnormal{F}$ by almost $50\%$ and, thus, only gives a very qualitative description of the density response at the third harmonic of the original perturbation. In addition, we find a complicated, nontrivial behaviour as $\chi^{(3)}(q)$ changes its sign around the Fermi wave number. We expect this to be a real physical effect, as a quite similar behaviour has been detected on the basis of our PIMC data at a lower density, $r_s=6$, shown in Fig.~\ref{fig:chi_rs6_panel} below.


\subsection{\label{sec:strong_coupling}Strong coupling: $r_s=6$}

\begin{figure*}\centering
\includegraphics[width=0.97\textwidth]{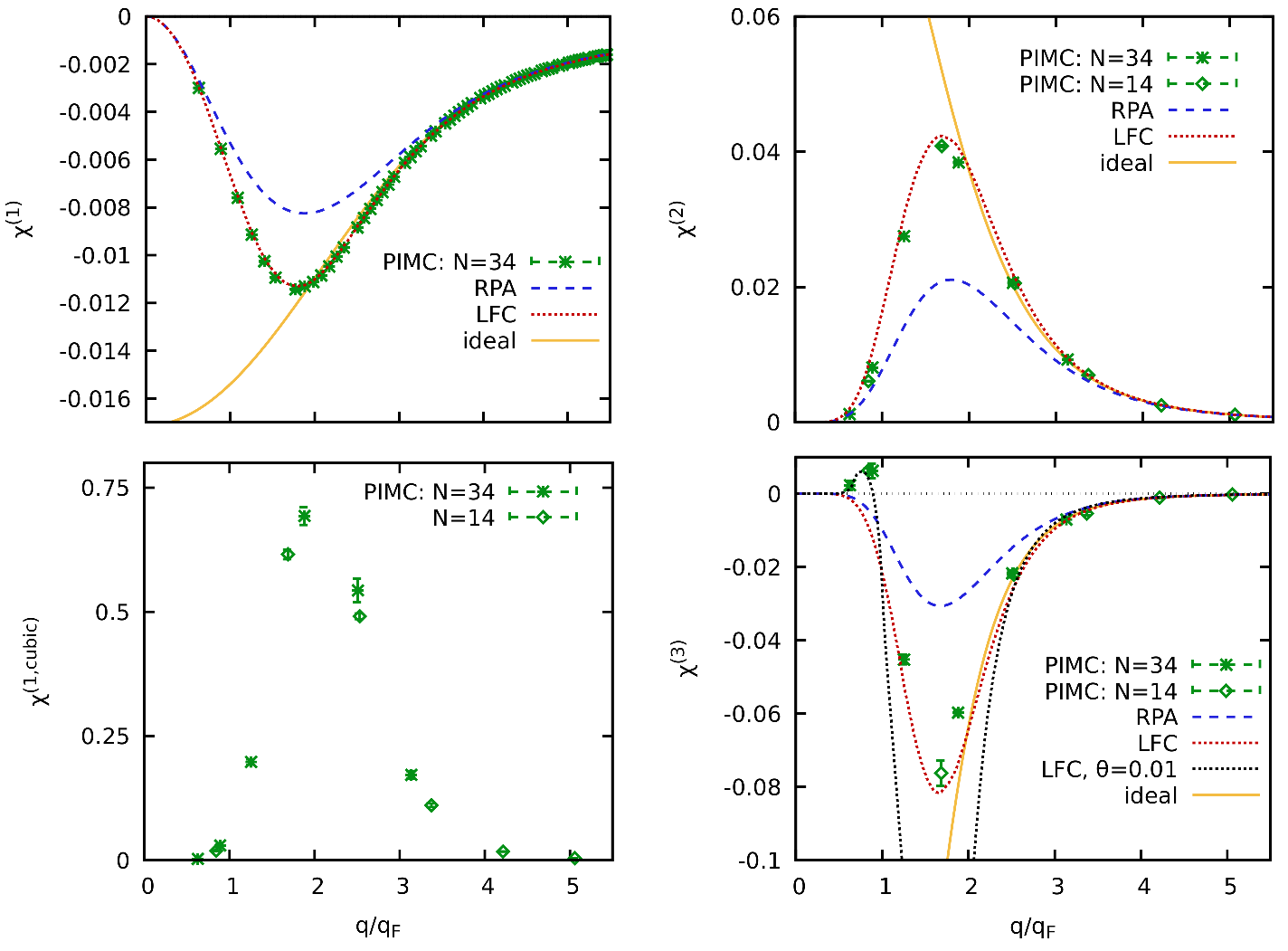}
\caption{\label{fig:chi_rs6_panel} Wave-number dependence of the linear response function $\chi^{(1)}(q)$ (top left), the quadratic response function of the second harmonic $\chi^{(2)}(q)$ (top right), the cubic response function of the first harmonic $\chi^{(1,cubic)}(q)$ (bottom left), and the cubic response function of the third harmonic $\chi^{(3)}(q)$ (bottom right). The green stars and diamonds correspond to PIMC data points for $N=34$ and $N=14$, the solid yellow curve to the ideal Fermi gas, the dashed blue curves to RPA, and the dotted red curves have been obtained using the LFC from Ref.~\cite{dornheim_ML}.
All data have been obtained for $r_s=6$ and $\theta=1$. }
\end{figure*}

Let us next investigate the nonlinear density response of the UEG at stronger coupling, $r_s=6$ and $\theta=1$. Such conditions can be realized experimentally for example in evaporation experiments~\cite{benage,karasiev_importance,low_density1,low_density2} and constitute a challenging benchmark for models and theories due to the pronounced impact of electronic exchange--correlation effects
\cite{Dornheim_PRL_2020_ESA,low_density1}.

Since the determination of the different generalized response functions from \textit{ab initio} PIMC data for $\braket{\hat\rho_\mathbf{k}}_{q,A}$ is completely analogous to the case of $r_s=2$ that was discussed in the previous sections, here we restrict ourselves to a concise discussion of the final results summarized in Fig.~\ref{fig:chi_rs6_panel}.

The top left panel corresponds to the usual LRT function $\chi^{(1)}(q)$, where the green stars have been obtained from a single simulation of the unperturbed UEG by evaluating Eq.~(\ref{eq:static_chi}) for $N=34$ unpolarized electrons. These data are in good agreement with the dotted red curve that has been computed from the static LFC presented in Ref.~\cite{dornheim_ML}, whereas RPA (dashed blue) substantially underestimates the magnitude of the minimum of $\chi^{(1)}(q)$ around $q=2q_\textnormal{F}$ by more than $25\%$. This is of course expected and a direct consequence of the increased importance of electronic correlation effects at low density~\cite{dornheim_physrep_18}. The solid yellow curve depicts results for the ideal Fermi gas and becomes increasingly inaccurate towards small wave-numbers, when screening effects start to manifest in the UEG~\cite{kugler_bounds}.

The top right panel shows the wave-number dependence of the quadratic response function of the second harmonic, $\chi^{(2)}(q)$. We find that this function exhibits a significantly faster decay towards zero both in the small- and large-$q$ limits, similar to our findings for $r_s=2$ above. In addition, the electronic LFC $G(q)$ has a substantially larger impact on $\chi^{(2)}(q)$ compared to $\chi^{(1)}(q)$, and the RPA underestimates the peak at $q\approx1.75q_\textnormal{F}$ by $50\%$. In contrast, the LFC curve is in good agreement with the PIMC data points for both $N=14$ (green diamonds) and $N=34$ (green stars) over the entire depicted wave-number range, and the deviation due to the approximate treatment of screening effects [cf.~Sec.~\ref{sec:theory}] is small. In addition, the PIMC data points for the two system sizes exhibit a smooth progression, which means that no finite-size effects can be resolved within the given level of accuracy. This is in good agreement with the recent investigation of $\chi^{(1,\textnormal{cubic})}(q)$ presented in Ref.~\cite{Dornheim_PRL_2020}.

Let us next focus on the bottom left panel showing the cubic density response function of the first harmonic, $\chi^{(1,\textnormal{cubic})}(q)$ at the same conditions. In particular, this function is sharply peaked around $q\approx2q_\textnormal{F}$ and quickly vanishes both in the limits of small and large wave numbers. We again mention that the evaluation of the RPA and LFC expressions given in Eqs.~(\ref{eq:cubic_first_RPA}) and (\ref{eq:cubic_first_LFC}) is not possible as $\chi_0^{(1,\textnormal{cubic})}(q)$ cannot be readily evaluated.

Finally, we investigate the cubic density response function of the third harmonic, $\chi^{(3)}(q)$ shown in the bottom right panel of Fig.~\ref{fig:chi_rs6_panel}. While the associated density response at the third harmonic is relatively small and unimportant compared to the signals at $\mathbf{k}=\mathbf{q}$ and $\mathbf{k}=2\mathbf{q}$, this function still deserves close attention as it exhibits interesting, nontrivial behaviour. More specifically, the PIMC data points are positive for small wave numbers and $\chi^{(2)}(q)$ changes its sign around the Fermi wave number, $q=q_\textnormal{F}$.

\begin{figure}\centering
\includegraphics[width=0.485\textwidth]{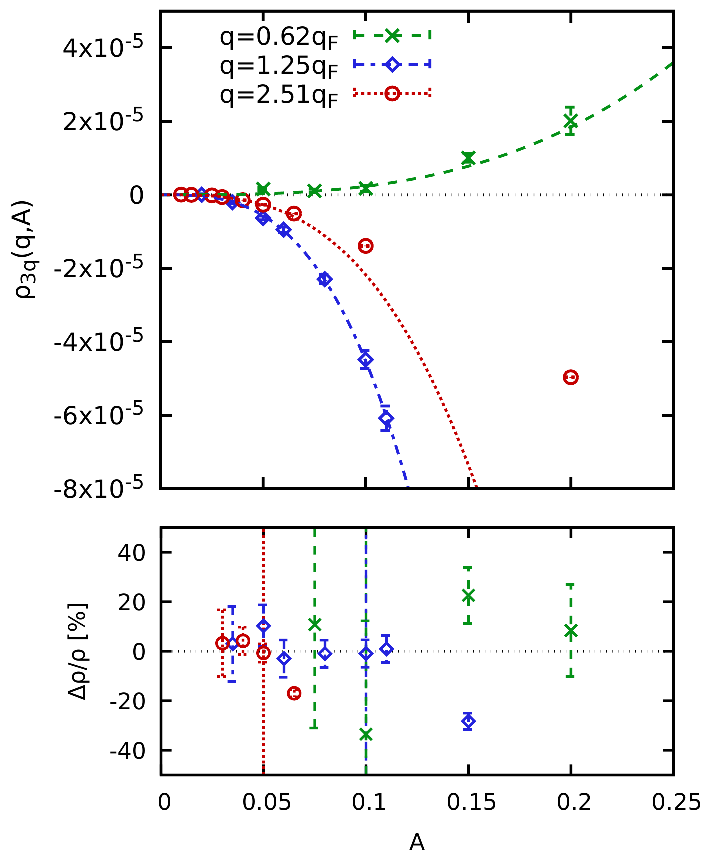}
\caption{\label{fig:determine_third_rs6} Determination of the cubic response function $\chi^{(3)}(q)$.
Top panel: density response of the third harmonic $\braket{\hat\rho_{3\mathbf{q}}}_{q,A}$ of the UEG for $N=34$, $r_s=6$, and $\theta=1$ for $q\approx0.62q_\textnormal{F}$ (green crosses) $q\approx1.25q_\textnormal{F}$ (blue diamonds), and $q\approx2.51q_\textnormal{F}$ (red circles). The corresponding curves of the same colour have been obtained by fitting Eq.~(\ref{eq:third_fit}) to the PIMC data. Bottom panel: relative deviation of the fits to the PIMC data. The vertical lines show the respective maximum value of $A$ that was included in the fits.
}
\end{figure}

This trend is further substantiated in Fig.~\ref{fig:determine_third_rs6}, where we show the dependence of the actual density response at the third harmonic $\braket{\hat\rho_{3\mathbf{q}}}_{q,A}$ on the perturbation amplitude $A$. The green crosses show results for the smallest wave number accessible for $N=34$ electrons, $q\approx0.62q_\textnormal{F}$, and we can clearly resolve a positive density response that monotonically increases with $A$.
In contrast, the blue diamonds and red circles corresponding to two larger wave numbers ($q\approx1.25q_\textnormal{F}$ and $q\approx2.51q_\textnormal{F}$, respectively) both exhibit a negative density response. 
As usual, the different curves have been obtained by fitting Eq.~(\ref{eq:third_fit}) to the PIMC data, and the bottom panel of Fig.~\ref{fig:determine_third_rs6} shows the relative deviation. Interestingly, the fit function remains accurate for larger $A$ at both $q=0.62q_\textnormal{F}$ and $q=1.25q_\textnormal{F}$ than for $q=2.51q_\textnormal{F}$, even though the response is substantially larger at the intermediate wave number (see also the vertical lines in the bottom panel showing the maximum value of $A$ that has been included into the fitting procedure).

Let us get back to the full wave-number dependence of $\chi^{(3)}(q)$ shown in the bottom right panel of Fig.~\ref{fig:chi_rs6_panel}. Evidently, neglecting electronic exchange--correlation effects within the RPA (dashed blue curve) results in a very poor description of the cubic response function, and the real depth of the minimum is underestimated by a factor of two. Including the static LFC gives the dotted red curve that is in very good agreement with the PIMC data points for $q\gtrsim q_\textnormal{F}$. 
Yet, the LFC result, Eq.~(\ref{eq:chi3_LFC}), is not capable to describe the sign change in $\chi^{(3)}(q)$ around the Fermi wave number and, thus, stays negative over the entire $q$-range.
This, in turn, means that the approximate linear treatment of screening effects [cf.~Sec.~\ref{sec:theory} above] leads to a description where this sign change is missed.

Finally, the dotted black curve has also been obtained by evaluating Eq.~(\ref{eq:chi3_LFC}) including the LFC, but for $\theta=0.01$, i.e, in the ground state. While this leads to a curve with a substantially deeper minimum around $q\sim1.5q_\textnormal{F}$ than at the Fermi temperature, it does exhibit the sign change in $\chi^{(3)}(q)$ that we find in our PIMC data at the higher temperature.

\section{Summary and Discussion\label{sec:summary}}

In this work, we have presented an in-depth analysis of nonlinear effects in the density response of the warm dense electron gas, substantially extending the first investigation of Ref.~\cite{Dornheim_PRL_2020}.
We have obtained extensive new PIMC results 
carrying out simulations of a harmonically perturbed electron gas, which has allowed us to obtain the full spectrum of excitations at the integer harmonics of the original perturbation. First and foremost, we have found that the dominant nonlinear contribution is given by the quadratic response at the second harmonic, for weak to moderate values of the perturbation amplitude $A$. The second potentially important nonlinear term is given by the cubic response at the first harmonic studied in Ref.~\cite{Dornheim_PRL_2020}, whereas the cubic response at the third harmonic is practically negligible in most realistic situations.
In addition, we have found that the nonlinear response functions more strongly depend on system parameters such as the temperature compared to the usual LRT, which makes them a potentially valuable new tool of diagnostics for WDM experiments. In addition, the nonlinear response is strongly shaped by electronic exchange--correlation effects, which makes a reliable theory based on \textit{ab initio} PIMC data indispensable. 
Furthermore,
we have extended previous works on the nonlinear density response of warm dense electrons and presented new expressions for the response functions
in terms of the well-known LFC that is readily available based on previous PIMC simulations~\cite{dornheim_ML,Dornheim_PRL_2020_ESA,Dornheim_PRB_2021}. In particular, this approach is capable to reproduce the benchmark PIMC data for the nonlinear density response with 
negligible 
computational cost. 

While the present analysis was based on simulations for a purely static density perturbation, even stronger nonlinear effects are to be expected for dynamic excitations. For example, the excitation of warm dense matter by a strong laser field will give rise to a time-dependent density response that will involve harmonics. If proper resonance conditions are fulfilled, as e.g. in the case of high harmonics generation in rare gases \cite{schafer_prl_93}, or laser plasmas \cite{haberland_01_pre}, the excitation of even higher harmonic orders should be expected.

We are convinced that the findings of our paper will open up many new avenues of future research in multiple directions. For example, a realistic perturbation in a WDM experiment will be given by a superposition of many distinct harmonic perturbations. While this has no profound consequence within LRT, a nonlinear treatment of the density response will give rise to mode-coupling between different perturbations, which will make the excitation spectrum more interesting and complicated. One can imagine two-color x-ray beam experiments to intentionally use the mode coupling for diagnostics. A second topic for future research is given by the investigation of the dynamic density response, which can be done on the basis of our new theoretical expressions and a suitable local field correction such as the ESA presented in Refs.~\cite{Dornheim_PRL_2020_ESA,Dornheim_PRB_2021}.
In addition to these basic questions, we stress that an accurate theory of nonlinear effects can be directly used for the improved computation of different material properties such as the electronic stopping power, screened potentials, or energy relaxation rates.


\section*{Acknowledgments}
This work was partly funded by the Center for Advanced Systems Understanding (CASUS) which is financed by Germany's Federal Ministry of Education and Research (BMBF) and by the Saxon Ministry for Science, Culture and Tourism (SMWK) with tax funds on the basis of the budget approved by the Saxon State Parliament, and by the Deutsche Forschungsgemeinschaft (DFG) via project BO1366/15.
The PIMC calculations were carried out at the Norddeutscher Verbund f\"ur Hoch- und H\"ochstleistungsrechnen (HLRN) under grant shp00026, on a Bull Cluster at the Center for Information Services and High Performace Computing (ZIH) at Technische Universit\"at Dresden,
on the cluster \emph{hemera} at Helmholtz-Zentrum Dresden-Rossendorf (HZDR), and at the computing center (Rechenzentrum) of Kiel university.

\bibliography{bibliography,mb-ref}
\end{document}